\documentclass{article}

\PassOptionsToPackage{numbers, compress}{natbib}

\usepackage[final]{neurips_2024}

\usepackage[utf8]{inputenc} 
\usepackage[T1]{fontenc}    
\usepackage{hyperref}       
\usepackage{url}            
\usepackage{booktabs}       
\usepackage{amsfonts}       
\usepackage{nicefrac}       
\usepackage{microtype}      
\usepackage{xcolor}         

\usepackage{graphicx}
\usepackage{mathtools}
\usepackage{amsmath}
\usepackage{wrapfig}
\usepackage{adjustbox}
\usepackage{multicol}
\usepackage{multirow}
\usepackage{wrapfig}
\usepackage{bm}
\usepackage[linesnumbered,ruled,vlined]{algorithm2e}
\usepackage{subcaption}

\newenvironment{packeditemize}{
\begin{list}{$\bullet$}{
\setlength{\labelwidth}{8pt}
\setlength{\itemsep}{0pt}
\setlength{\leftmargin}{\labelwidth}
\addtolength{\leftmargin}{\labelsep}
\setlength{\parindent}{0pt}
\setlength{\listparindent}{\parindent}
\setlength{\parsep}{0pt}
\setlength{\topsep}{1pt}}}{\end{list}}

\usepackage[accsupp]{axessibility}  

\usepackage{orcidlink}
\newcommand{\sysname}{\texttt{COSMIC}\xspace}
\renewcommand\footnotemark{}

\begin{document}

\title{COSMIC: Compress Satellite Images Efficiently via Diffusion Compensation}

\author{
Ziyuan Zhang$^1$\quad  Han Qiu$^{1}$*\thanks{*Corresponding authors.} \quad Maosen Zhang$^1$ \quad Jun Liu$^1$*\\ 
\textbf{Bin Chen$^2$ \quad Tianwei Zhang$^3$\quad Hewu Li$^1$}\\
$^1$ Tsinghua University, China \\
$^2$ Harbin Institute of Technology, Shenzhen, China \\
$^3$ Nanyang Technological University, Singapore\\
\tt\footnotesize{\{ziyuan-z23,zhangms24\}@mails.tsinghua.edu.cn}, \tt\footnotesize{\{qiuhan,juneliu\}@tsinghua.edu.cn} \\ 
\tt\footnotesize{chenbin2021@hit.edu.cn}, \tt\footnotesize{tianwei.zhang@ntu.edu.sg}, 
\tt\footnotesize{lihewu@cernet.edu.cn}
}

\maketitle

\begin{abstract}

With the rapidly increasing number of satellites in space and their enhanced capabilities, the amount of earth observation images collected by satellites is exceeding the transmission limits of satellite-to-ground links. 
Although existing learned image compression solutions achieve remarkable performance by using a sophisticated encoder to extract fruitful features as compression and using a decoder to reconstruct, 
it is still hard to directly deploy those complex encoders on current satellites' embedded GPUs with limited computing capability and power supply to compress images in orbit. 
In this paper, we propose \sysname, a simple yet effective learned compression solution to transmit satellite images. 
We first design a lightweight encoder (i.e. reducing FLOPs by $2.6\sim 5\times $) on satellite to achieve a high image compression ratio to save satellite-to-ground links. 
Then, for reconstructions on the ground, to deal with the feature extraction ability degradation due to simplifying encoders, we propose a diffusion-based model to compensate image details when decoding. 
Our insight is that \textit{satellite's earth observation photos are not just images but indeed multi-modal data with a nature of Text-to-Image pairing} since they are collected with rich sensor data (e.g. coordinates, timestamp, etc.) that can be used as the condition for diffusion generation.  
Extensive experiments show that \sysname outperforms state-of-the-art baselines on both perceptual and distortion metrics.
The code is publicly available at \url{https://github.com/Joanna-0421/COSMIC}.

\end{abstract}

\section{Introduction}
\label{sec:intro}

The revival of the aerospace industry~\cite{dreyer2009latest,martin2018bolstering}, coupled with reduced costs of launching rockets~\cite{frick2018small}, has fueled an exponential increase in the number of nanosatellites, resulting in massive growth in images collected in-orbit. 
For example, the Sentinel-3 missions can collect a maximum of 20 TB raw data on satellites (mainly earth observation images) every day~\cite{esch2018exploiting}.
However, the data transmission capability between satellites and ground stations has clear upper bounds~\cite{devaraj2017dove,tao2024known,denby2023kodan}. 
This situation of the rapid growth of images collected by satellites versus the limited transmission capability to the ground requires effective image compression on satellites before transmission back to Earth.

\begin{wrapfigure}{r}{0.5\textwidth}
    \centering 
    \includegraphics[width=0.5\textwidth]{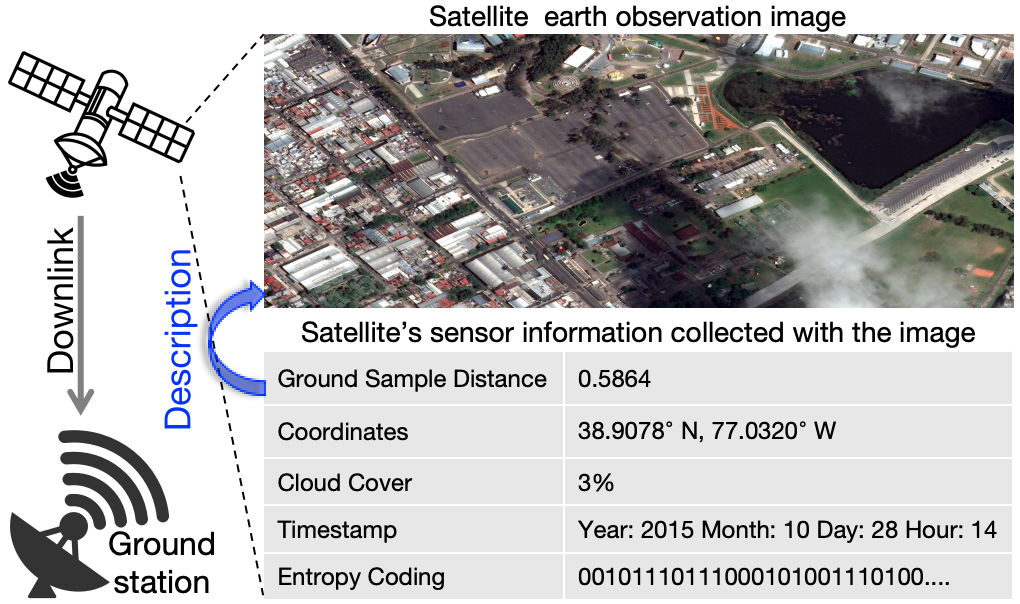}
    \caption{An example of the satellite's earth observation image and this image's corresponding sensor data as a description.}
    \label{fig:simg_pair}
\end{wrapfigure}

Current industrial compression solutions for satellite images rely on JPEG~\cite{wallace1991jpeg}, JPEG2000~\cite{taubman2002jpeg2000}, or CCSDS123~\cite{hernandez2021ccsds} (e.g. satellite BilSAT-1~\cite{yuksel2005bilsat}). 
These solutions are outperformed by various learned compression methods~\cite{cheng2020learned} in various cases.  
Existing learned image compression methods~\cite{mentzer2020high,li2022content,yang2024lossy} use sophisticated encoders to extract fruitful features and then use a decoder to decompress~\cite{johnston2019computationally,yang2023computationally}. 
Although we notice a novel promising trend of \textit{deploying embedded GPUs on satellites} in both academia~\cite{denby2020orbital, denby2023kodan, adams2019towards, tao2024known} and industry (e.g. satellite Phi-Sat-1~\cite{giuffrida2021varphi}, Chaohu-1~\cite{StarDetect}, and Forest-1~\cite{OroraTech}) which brings the potential opportunity of using learned compressors on satellites.  
It is still hard to directly adopt existing learned compression solutions for satellites since their sophisticated encoders are still too complex for GPUs on satellites (e.g. NVIDIA Jetson Xavier NX on Forest-1~\cite{OroraTech}) which have limited computing capacity and power supply~\cite{jetson-xavier-nx}. 
We have two insights to fill the above gaps. 
(1) We first design a lightweight encoder on satellites with a higher priority of compression ratio than feature extraction ability.
(2) At the receiver's end on the ground, we deal with this simple encoder's feature extraction ability degradation by compensating image contents when decoding. 
We choose diffusion as compensation due to its powerful generation capability and, more importantly, \textit{satellites' earth observation photos are not only images but enjoy a multi-modal nature in which rich real-time sensor information is the description of the corresponding photo}. 
For instance, in~\figurename~\ref{fig:simg_pair}, the coordinates (e.g. latitude and longitude) denote the location of the image which describes its main category (e.g. sea, city, etc.) and the timestamp describes the image's lightning-like day or night. 

In this paper, we propose to \textbf{CO}mpress \textbf{S}atellite i\textbf{M}age via d\textbf{I}ffusion \textbf{C}ompensation  (\sysname), a novel learned image compression method for satellites. 
\sysname has two key components, i.e., (1) a lightweight encoder for compression on satellite and (2) a sophisticated decompression process with a decoder and a diffusion model on the ground with sufficient GPUs.
First, we design a lightweight convolution architecture to extract local features and apply convolution to obtain an attention map of global features, to realize a lightweight image compression encoder in terms of FLOPs. 
Please note that lightweight encoders usually extract fewer key features which increases the difficulty for the decoder to decompress. 
Thus, our second component, decompression, has two parts including decoding with a corresponding decoder and, more importantly, a compensation model.  
Inspired by the multi-modal nature of the satellite's images, we aim to build text-to-image pairs (image and its sensor information like~\figurename~\ref{fig:simg_pair}) and use diffusion as the compensation model.

We compared \sysname with 6 state-of-the-art (SOTA) baselines, 3 of which are based on generative models, considering both distortion and perceptual metrics.
In addition, we constructed two image compression test sets based on satellite images by considering ordinary scenes and unique tile scenes in satellite imagery. 
Extensive experiments have proven that \sysname significantly reduces the encoder's complexity to $2.6\sim 5\times $ fewer FLOPs while achieving better performance on almost all metrics than baselines. 
Our contributions can be summarized as follows.
\begin{packeditemize}
    \item We propose a novel idea that uses a lightweight image compression encoder on satellites and leverages satellite images' text-to-image pairing nature for compensation when decompressing.  
    
    \item We propose a novel compensation model based on stable diffusion to compensate image details when decompressing with the unique sensor data of satellite images as descriptions.
    
    \item We analyzed the characteristics of satellite images in detail and incorporated them into the training and inference stages. In addition, we constructed two datasets under satellite image transmission scenarios, taking into account the typical satellite image transmission tasks like tile scenes. 
\end{packeditemize}

\section{Background}
\subsection{Earth observation missions on satellites}

Earth observation missions (e.g. NASA's Landsat Program~\cite{wulder2022fifty}) involve the use of satellite photos to monitor and collect data for tasks like forestry~\cite{achard2010estimating}, agriculture~\cite{crocetti2020earth,petropoulos2018earth}, land degradation~\cite{de2011quantitative}, land use and land cover~\cite{pandey2021land}, biodiversity~\cite{kuenzer2014earth}, and water resource~\cite{lawford2013earth, uereyen2019review}.
Traditional earth observation missions rely on a pipeline in which satellites take photos and then send them back to ground stations for analysis. 
Recently, along with the reduced cost of launching rockets and manufacturing nanosatellites~\cite{esposito2019highly,adams2019towards}, the photos on satellites are rapidly increasing which brings novel challenges for transmitting photos back to the ground. 
A recent promising approach is to deploy embedded GPUs on satellites to support DNN models to either filter useless data before transmission or make partial processing tasks on satellites. 
For instance, ESA's satellite Phi-Sat-1~\cite{giuffrida2021varphi} first deploy Intel VPU on satellite to support DNN models for filtering useless photos (e.g. covered by clouds) that can save 30\%+ transmission volume. 
OroraTech has launched AI nanosatellites with the NVIDIA Jetson Xavier NX for wildfire detection~\cite{OroraTech}, and Orbital Sidekick uses NVIDIA Jetson AGX Xavier as the AI engine at the edge of the satellite to detect gas pipeline leaks~\cite{Orbital-Sidekick}. 
However, due to the inelastic computational capabilities of onboard satellites and limited power supply only from the sunshine (i.e., up to 15 Watt for GPUs on satellites~\cite{OroraTech}), a certain amount of images are still needed to be transmitted back to the ground. 
This brings an urgent need for satellite-specific image compression methods.

\subsection{Neural image compression methods}
\label{LIC_background}

\noindent\textbf{Learned image compression methods} have achieved remarkable rate-distortion performance compared with classical information theory-based image compression methods, 
to each of which consists of an encoder $\mathcal{E}$, a quantization $\mathcal{Q}$, and a decoder $\mathcal{D}$.
The encoder, as the most critical part, extracts key features from the image as the latent representation. 
Higher quality representation extracted by the encoder means less content loss at compressing which is more likely to reconstruct a higher quality image when decompression. 
Thus, SOTA approaches explore introducing more complex modules into the encoder.
\cite{liu2023learned} integrates the transformer into the CNN encoder and uses the transformer-CNN mixture block to extract rich global features. 
The other approaches aims to reduce the complexity of the decoders that are deployed on edge devices like smartphones. For instance, 
\cite{yang2023computationally} adopts shallow or even linear decoding transforms to reduce the decoding complexity, compensated by more powerful encoder networks and iterative encoding.

\noindent\textbf{Generative models for decompression.}
Although VAE-based methods have achieved good performances, optimizing solely for mean square error (MSE) can lead to excessive image smoothing, resulting in visual artifacts.  
More recent works~\cite{li2022content, yang2024lossy, hoogeboom2023high, ghouse2022neural} have combined VAEs with generative models (e.g. diffusion) to achieve better visual results. 
\cite{yang2022lossy} uses a conditional diffusion model as the image compression decoder which improves visual results. 
\cite{hoogeboom2023high} and \cite{ghouse2022neural} decouple the compression task and augmentation task, sending the output of the VAE codec to diffusion to predict the residual.

\noindent\textbf{Satellite image compression method.}
There are some compression methods specifically for remote sensing images~\cite{zhang2023global, fu2023remote, xiang2024remote}.
~\cite{xiang2024remote} uses discrete wavelet transform to divide image features into high-frequency features and low-frequency features, and design a frequency domain
encoding-decoding module to preserve high-frequency information, thereby improving the compression performance.
\cite{fu2023remote} explore local and non-local redundancy through a mixed hyperprior network to improve entropy model estimation accuracy.
Few of these works focus on onboard deployment. 
\cite{guerrisi2023artificial} use the CAE model to extract image features and reduce the image dimension to achieve compression, and deploy the model on VPU. 
However, this method only considers the reduction of image dimension and does not consider the arithmetic coding process in the actual transmission process, resulting in the image compression rate can only be adjusted by changing the model architecture.

\noindent\textbf{Limitations to use for satellites.} Most of the above approaches don't consider lightweight compression encoders which makes them impractical to deploy on satellite's embedded GPUs constrained by computing capacity and power supply. 
The approach used on VPU is impractical as the image compression rate is highly related to model architecture.
Besides, none of them pay attention to the multi-modal nature of satellite earth observation images to introduce conditions to further improve decompression quality.

\section{Prelimiaries}

\subsection{Problem formulation}
\label{sec:problem_formulation}

We formulate the basic process of learned image compression. 
The encoder $\mathcal{E}$ uses a non-linearly transformation to convert the input image $\mathbf{x}$ into the latent representation $\mathbf{y}$, which is subsequently discretized and entropy-coded by quantization $\mathcal{Q}$ under a learned hyper prior $\mathrm{\zeta}$.
Under the stochastic Gaussian model, each discrete code ${\lfloor \mathbf{y} \rceil}_\mathrm{i}$ can be expressed as a Gaussian distribution with mean $\mathrm{\mu} _\mathrm{i}$ and variance $\mathrm{\sigma} _\mathrm{i}$ given a hyper prior $\mathrm{\zeta} _\mathrm{i}$: $\mathrm{p}\left( {\lfloor \mathbf{y} \rceil}_\mathrm{i}|\mathrm{\zeta} _\mathrm{i} \right) =\mathcal{N} \left( \mathrm{\mu} _\mathrm{i},\mathrm{\sigma} _\mathrm{i}^{2} \right)$.
The decoder $\mathcal{D}$ reconstructs the discrete representation $\lfloor \mathbf{y} \rceil$ to the image $\hat{\mathbf{x}}$.
The model can be optimized by the loss function (Eq.~\ref{eq:loss_ic}).
\begin{equation}
\label{eq:loss_ic}
    \mathcal{L} _{\mathrm{IC}}=\mathrm{R}+\mathrm{\lambda D}=\mathbb{E} \left[ -\log _2\mathrm{p}\left( \lfloor \mathbf{y} \rceil|\mathrm{\zeta} \right) -\log _2\mathrm{p}\left( \mathrm{\zeta} \right) \right] +\mathrm{\lambda}\mathbb{E} \left[ \mathrm{d}\left( \mathbf{x},\hat{\mathbf{x}} \right) \right], 
\end{equation}
where $\mathrm{R}$ is the bit rate of latent discrete coding, $\mathrm{D}$ is the distortion between the original and the reconstructed image (measured by MSE), and $\lambda$ controls the trade-off between rate and distortion.

\subsection{Diffusion model}

Diffusion model is a type of generative model that can generate images from Gaussian noise through multi-step iterative denoising.
These models include two Markov processes.
First, the diffusion process gradually applies noise to the image until the image is destroyed and becomes complete Gaussian noise. 
Then, in the reverse stage, it learns the process of restoring the Gaussian noise to the original image. 
During the inference stage, given a random noise sample $\mathbf{x}_\mathrm{T}\sim \mathcal{N} \left( 0,1 \right) $, the diffusion model can denoise through T steps and gradually generate a photorealistic image $\mathbf{x}_0$. 
At each step $\mathrm{t}\in \left\{ 0,1,...,\mathrm{T} \right\}$, intermediate variable $\mathbf{x}_\mathrm{t}$ can be expressed as $\mathbf{x}_{\mathrm{t}}=\sqrt{1-\mathrm{\beta}_{\mathrm{t}}}\mathbf{x}_{\mathrm{t}-1}+\mathrm{\beta}_{\mathrm{t}}{\bm{\epsilon}}_{\mathrm{t}}$, 
where $\mathrm{\beta} _\mathrm{t}\in \left( 0,1 \right) $ is the variance hyperparameter of Gaussian distribution, and satisfies $\mathrm{\beta} _1<\mathrm{\beta} _2<...<\mathrm{\beta} _\mathrm{T}$; $\bm{\epsilon} _\mathrm{t}\sim \mathcal{N} \left( 0,1 \right) $ is the Gaussian noise at step $\mathrm{t}$. 
In the diffusion model, a noise prediction network ($\bm{\epsilon} _{\mathrm{\theta}}$) is used to predict the noise at step $\mathrm{t}$, and $\mathbf{x}_\mathrm{t-1}$ can be obtained from $\mathbf{x}_\mathrm{t}$. 

The diffusion model can also understand the content of the given conditions, such as text and images, and generate images consistent with the conditions.
In this case, the noise prediction network takes three parameters: intermediate sample $\mathbf{x}_\mathrm{t}$, timestep $\mathrm{t}$, and given condition $\mathrm{\varsigma} $ as input.
To guarantee the noise predicted by the noise prediction network at the $\mathrm{t}$-th step of the reverse process has the same distribution as the noise introduced into the image during the diffusion process, diffusion model usually use $\mathcal{L} _2$ to optimize the network following $\nabla _{\mathbf{\theta}}\left\| \bm{\epsilon}_{\mathrm{t}}-\bm{\epsilon}_{\mathrm{\theta}}\left( \mathbf{x}_{\mathrm{t}},\mathrm{t},\mathrm{\varsigma} \right) \right\|$.

Stable diffusion~\cite{rombach2022high} is proposed to reduce the training cost, which implements the diffusion process in a low-dimensional latent space while retaining the high-dimensional information in the original pixel space for decoding. 
In this article, we aim to leverage the powerful generation ability of stable diffusion and its ability to maintain consistency with a given condition to provide compensation for the information lost by the image compression encoder.

\section{Method}

\subsection{Framework of \sysname}

\begin{figure}[htbp]
    \centering
    \includegraphics[width=0.95\linewidth]{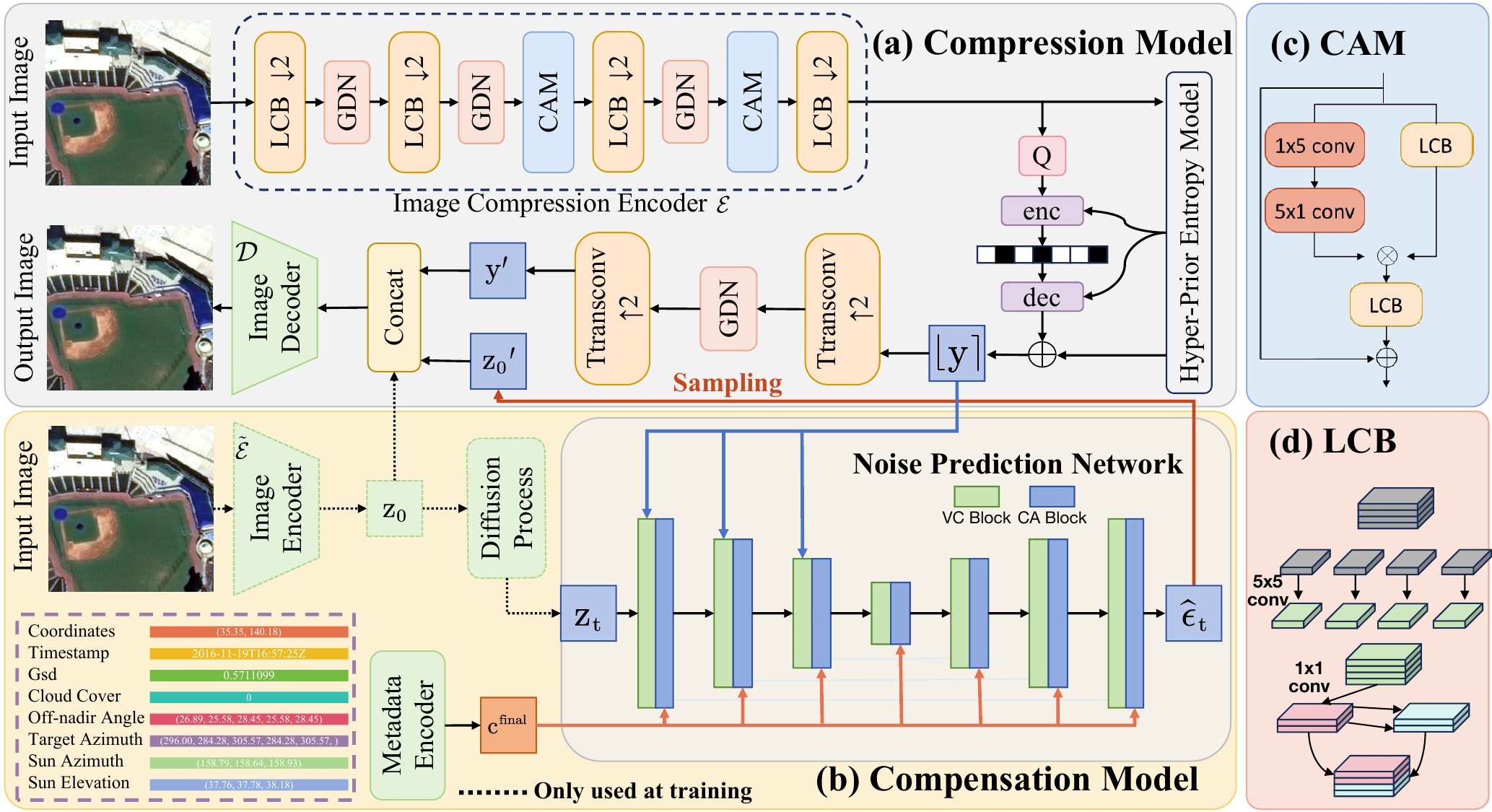}
    \caption{\textbf{\sysname framework.} (a) Compression module for satellite images: a lightweight encoder and a compensation-based decoder (Sec.~\ref{sec:C_GIC}). (b) In the noise prediction network, each Cross-Attention (CA) block receives embedding of the Metadata Encoder (ME) (Sec.~\ref{sec:DC}), and the Vanilla Convolution (VC) blocks use latent image discrete encoding to guide the prediction of noise for each diffusion step. (c) \& (d) Convolution attention module and lightweight convolution block (Sec.~\ref{sec:LICE}).}
    \label{fig:overview}
\end{figure}

As illustrated in \figurename~\ref{fig:overview}, \sysname consists of two components: a compression module and a compensation module. 
To adapt to the satellite scenarios, the compression module includes a lightweight image compression encoder $\mathcal{E}$ and an entropy model deployed on the satellite (Sec.~\ref{sec:LICE}), as well as an image compression decoder $\mathcal{D}$ deployed at the ground station. 
To fix the content detail loss caused by the lightweight encoder, a compensation module is proposed, which is entirely deployed at the ground station.
It has an encoder $\tilde{\mathcal{E}}$, aiming to extract compensation information $\mathrm{z}_0$ from original images, which is received by decoder $\mathcal{D}$ as compensation for latent representation $\mathbf{y}\prime$ extracted by $\mathcal{E}$ during training (Sec.~\ref{sec:C_GIC}).
During the inference phase, the noise prediction network generates compensation information $\mathrm{z}_0\prime$ from noise to simulate $\mathrm{z}_0$ as a compensation (Sec.~\ref{sec:DC}).

\subsection{Lightweight image compression encoder}
\label{sec:LICE}

To make the image compression encoder $\mathcal{E}$ practical on satellites, we first make it lightweight. 
The main idea is to reduce the amount of calculation required for image compression in terms of FLOPs.
We followed the classic architecture of image compression encoder~\cite{balle2018variational}, which is structured with downsampling convolutions with a stride of 2 and generalized divisive normalization (GDN)~\cite{balle2015density} arranged alternately (\figurename~\ref{fig:overview} (a)). 
Since GDN provides the best performance for image compression when the number of channels is 192~\cite{balle2018efficient}, the increase in convolution filters will exponentially increase the calculation amount of convolution.
To reduce the amount of computation required for image downsampling, we propose a lightweight convolution block (LCB), as in \figurename~\ref{fig:overview} (d), which uses depthwise convolution to replace the ordinary convolution with a convolution kernel size of $5\times5$.
To interact between different channels of the feature map, depthwise convolution is followed by a $1\times1$ convolution with a full number of channels.
Inspired by~\cite{han2020ghostnet}, which proves that there is a lot of redundancy in the features extracted by convolution, in LCB, only half of the output feature maps are obtained through $1\times1$ convolution, and the remaining half of the output feature maps are obtained through linear transformations with cheap cost using redundant features.

Transformer-based methods~\cite{zou2022devil, liu2023learned} can outperform CNN-based methods as the attention can capture non-local information of the images. 
However, the computational complexity of self-attention has a quadratic relationship with the size of the input feature map, which is not computationally friendly.
Inspired by~\cite{tang2022ghostnetv2}, we use two one-dimensional convolutions in series.
The first convolution is used to extract horizontal information. 
On this basis, the second convolution is used to vertically synthesize the previously extracted horizontal information to obtain the global attention map.
Meanwhile, the other branch uses LCB with one stride to capture local information, as shown in \figurename~\ref{fig:overview} (c).

\subsection{Compensation-guided image compression}
\label{sec:C_GIC}

After giving a lightweight design of the image compression encoder $\mathcal{E}$, we note that the representation ability of this encoder is inevitably degraded.  
Specifically, the features contained in the latent representation $\lfloor \mathbf{y} \rceil$ received by the ground station are not enough to let the decoder reconstruct a high-quality image. 
Thus, we explore compensation for the degradation in encoding.

It is well-known that stable diffusion has powerful generation capabilities for specified content from noise under the guidance of text information~\cite{rombach2022high}. 
The ground station can obtain not only the latent representation compressed by the encoder but also rich sensor data like the geographical location, time, camera parameters, etc. along with each image.  
We use these sensor data as conditions to guide diffusion generation to fix the missing image details. 
The training is divided into two stages. 
In the first stage, we train the compression model.
As shown in (\figurename~\ref{fig:overview} (b)), 
since the Image decoder $\mathcal{D}$ needs two parts of information (i.e. $\mathbf{y}\prime$ and $\mathbf{z_0}$ in \figurename~\ref{fig:overview}) for decoding, we introduce another image encoder $\tilde{\mathcal{E}}$ to extract compensation information $\mathbf{z_0}$ from the original image. 
In the first stage, $\mathcal{E}$, $\tilde{\mathcal{E}}$ and $\mathcal{D}$ are trained together.
The reconstructed image $\hat{\mathbf{x}}$ can be expressed as in Eq.~\ref{eq:reconstructed}. 
\begin{equation}\label{eq:reconstructed}
    \hat{\mathbf{x}}=\mathcal{D} \left( \mathrm{concat}\left( \mathrm{transconv}\left( \lfloor \mathbf{y} \rceil \right) ,\tilde{\mathcal{E}}\left( \mathbf{x}_0 \right) \right) \right) 
\end{equation}

In the second stage of training, we freeze the parameters of $\mathcal{E}$, $\tilde{\mathcal{E}}$ and $\mathcal{D}$, and train the noise prediction network, with the goal of making the information generated by the diffusion model as close to $\mathbf{z_0}$ as possible, denoted as $\mathbf{z_0}\prime$, so as to generate the compensation information required by the decoder.
During the inference phase, the trained diffusion model can generate compensation information $\mathbf{z_0}\prime$. 
Therefore, we no longer need $\tilde{\mathcal{E}}$. 
The $\mathbf{z_0}\prime$ generated by the diffusion model replaces the $\mathbf{z_0}$ extracted by $\tilde{\mathcal{E}}$ to help the image decoder decompress the image.

\subsection{Conditional diffusion model for loss compensation}
\label{sec:DC}

As pointed out in Sec.~\ref{sec:intro}, earth observation images collected by satellites are indeed multi-modal data. 
Here we consider that a satellite image $\mathbf{x}_0$,  when transmitted to the ground station, contains a discrete image coding $\lfloor \mathbf{y} \rceil$ paired with its sensor information denoted as numerical metadata $\mathrm{m}$. 
For $\mathrm{m}\in \mathbb{R} ^\mathrm{M}$, just like diffusion handles timestep $\mathrm{t}$, we use sinusoidal embedding ($\mathrm{E}_{\sin}$) to encode them to $\mathrm{c}_\mathrm{j}\in \mathbb{R} ^{1\times \mathrm{d}}\left( j=1,2,...\mathrm{M} \right)$, where $\mathrm{d}$ is the dimension of the clip embedding, as we concatenate the metadata embedding together as a description of an image.
This process is expressed as in Eq.~\ref{eq:eq5}. 
\begin{equation} \label{eq:eq5}
    \mathrm{c}^{\mathrm{final}}=\mathrm{MLP}\left( \mathrm{concat}\left( \left[ \mathrm{E}_{\sin}\left( \mathrm{m}_1 \right) ,...,\mathrm{E}_{\sin}\left( \mathrm{m}_M \right) \right] \right) \right)  
\end{equation}
This final metadata condition will be incorporated into the latent representation using CrossAttention (CA) blocks to guide the generation process.

Stable diffusion was originally used for generative tasks, which have randomness. 
Here, we expect that stable diffusion generates image details that are not extracted by the satellite image encoder $\mathcal{E}$, and still retain the overall structure of the image.
To address this problem, we inject the discrete image coding $\lfloor \mathbf{y} \rceil$ into the Vanilla Convolution (VC) blocks of the noise prediction network to provide the structure information and improve content consistency, which can be described in Eq.~\ref{eq:eq6}.
\begin{equation}\label{eq:eq6}
{\mathrm{f}_{\mathrm{i}}}^{\prime}=\mathrm{f}_{\mathrm{i}}+\mathrm{projection}_{\mathrm{i}}\left( \lfloor \mathbf{y} \rceil \right),
\end{equation}
where $\mathrm{f}_\mathrm{i}$ is the i-th feature map of the U-Net backbone, and $\mathrm{projection}_\mathrm{i}$ is the upsampling convolution used to align the dimensions between $\lfloor \mathbf{y} \rceil$ and $\mathrm{f}_\mathrm{i}$.
Guided by image coding and the metadata as a description, we use MSE loss to minimize the distance between target distribution and learned distribution in latent space as in Eq.~\ref{eq:eq7}.
\begin{equation}\label{eq:eq7}
    \mathcal{L} _{\mathrm{ldm}}=\mathbb{E} _{\mathrm{t},\mathrm{z}_0,\mathrm{\epsilon}\sim \mathcal{N} \left( 0,1 \right)}\left[ \left\| \bm{\epsilon}_{\mathrm{t}}-\bm{\epsilon}_{\mathrm{\theta}}\left( \mathrm{z}_{\mathrm{t}},\mathrm{t},\lfloor \mathbf{y} \rceil ,\mathrm{c}^{\mathrm{final}} \right) \right\| ^2 \right]   
\end{equation}

\section{Experiments}

\subsection{Setup}

\textbf{Dataset.} 
We use the function Map of the World (fMoW)~\cite{christie2018functional}, which has 62 categories, and in which each image is paired with different types of metadata features, as our training data and test data. 
For training data, we randomly crop the image to a resolution of $256\times256$ pixels.
For information collected by satellite sensors as metadata, we choose Coordinates, Timestamp, GSD, Cloud cover, Off-nadir Angle, Target Azimuth, Sun Azimuth, and Sun Elevation provided by fMoW.

For testing data, we constructed two test sets with different resolutions. 
One is from the fMoW test set, where to ensure a comprehensive representation of categories, we randomly selected one image from each category, cropped it to a resolution of $256\times256$, and used it as a standard test set. 
For another test set, we considered the actual scenario of the satellite to construct a tile test set. 
Since the image captured by the satellite is a large geographic region, the computing resources on the satellite are limited, and large-size images cannot be processed directly, so images should first be cut into smaller sub-images, this process is known as tiling~\cite{huang2018tiling, xu2022cloud, denby2023kodan,bosch2019semantic}.  
In this paper, for images of size $2306\times2306$, we first divide each image into 81 smaller patches of size $256\times256$ each, and then compress and decompress each patch individually, as illustrated in \figurename~\ref{fig:visual_test2} (a).
After obtaining all the decompressed patches, we reassemble them as one image for further evaluation.

\noindent \textbf{Metrics.}
We use 4 metrics for quantitative measures following previous works~\cite{mentzer2020high,li2022content,yang2024lossy}. 
For distortion comparison, we use the Peak Signal-to-Noise Ratio (PSNR) and Multi-Scale Structural Similarity Index Measure (MS-SSIM) to validate the pixel fidelity and measure brightness, contrast, and structural information at different scales. 
For perceptual comparison, we choose Learned Perceptual Image Patch Similarity (LPIPS) and Fréchet Inception Distance (FID). 

\noindent \textbf{Model training details.}
The training has two stages. 
First, we train the image compression encoder $\mathcal{E}$, image encoder $\tilde{\mathcal{E}}$ and image decoder $\mathcal{D}$ together using $\mathcal{L}_{\mathrm{IC}}$ for 100 epochs with a batchsize of 32.
Second, we freeze the parameters of the model trained in the first stage, use the pretrained stable diffusion model for the noise prediction network, and finetune it using $\mathcal{L}_{\mathrm{ldm}}$ for 10 epochs with a batchsize of 4.
All the training experiments are performed on $10\times$ NVIDIA GeForce RTX 3090 using Adam optimizer with $\mathrm{lr}=1\times 10^{-4}$ and $\lambda \in \left\{ 0.00067, 0.0013, 0.0026, 0.005 \right\}$.
During inference, we utilize the DDIM sampling~\cite{song2020denoising} with 25 steps.

\noindent \textbf{Baselines.}
We consider 6 baselines including traditional methods, VAE-based methods, and generative model based methods. 
\textbf{Elic}~\cite{he2022elic} proposes a multi-dimension entropy estimation model, which can effectively reduce the bit rate and improve the coding performance.
\textbf{Hific}~\cite{mentzer2020high} pays more attention to the perception of the model reconstruction effect, obtaining visually pleasing reconstructed images.
Based on Hific, \textbf{COLIC}~\cite{li2022content} considers the semantic information of the image when designing the loss function, and treats structure and texture respectively.
\textbf{CDC}~\cite{yang2024lossy} is the first work to use the diffusion model as an image compression decoder, performing the reverse process of diffusion in pixel space to reconstruct the image. 
\textbf{HL\_RS}~\cite{xiang2024remote} is an image compression method, especially for remote sensing images, which processes the high-frequency part and the low-frequency part of the images separately to better preserve the important high-frequency features of remote sensing images.
For these 5 baselines, we retrain their models with the fMoW dataset for a fair comparison.
Besides, we choose JPEG2000, the industrial solution for satellite image compression~\cite{yu2022analysis}, for comparison.

\subsection{Comparison with baselines}

\begin{figure}[!htbp]
    \centering
    \includegraphics[width=0.99\linewidth]{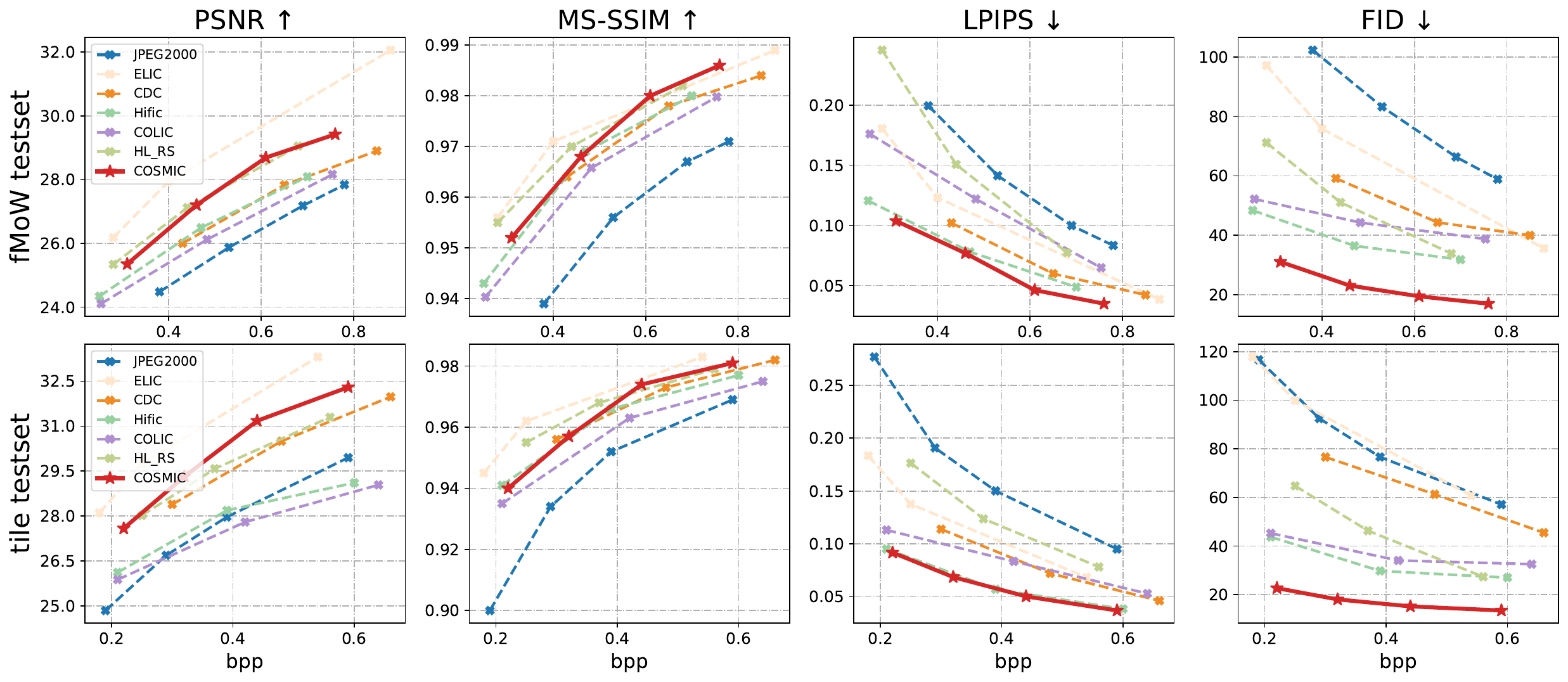}
    \caption{Trade-off between bitrates and different metrics on \sysname and baselines. The ↑ (↓) means higher (lower) is better. The first row is for the fMoW test set (image size $256\times256$). The second is for the tile test set by comparing between the stitched images and their original ones.}
    \label{fig:baseline_compare}
\end{figure}

\noindent\textbf{RD performance.}
\figurename~\ref{fig:baseline_compare} shows the comparison results with baselines on two test sets.
The dotted line in the figure represents the baselines, and the solid line represents \sysname.
We demonstrate results from two perspectives, i.e., distortion and perception. 
Across all transmission rates, \sysname surpasses the baselines in terms of LPIPS, and FID.
At low bpp, the MS-SSIM of \sysname is lower compared to the baseline.
This is mainly because as the bpp decreases, the encoder extracts less information, and during the decompression process, there is a greater reliance on diffusion-based generation (more details in Sec.~\ref{sec:ablation}). 
Additionally, due to the degraded feature extraction of the lightweight encoder, the feature obtained at low bpp is insufficient to guide the diffusion process in generating high-fidelity images. 
As the bpp increases, the latent coding contains more features, resulting in a significant improvement in the MS-SSIM of \sysname, demonstrating SOTA performance.
Note that the sensor data is transmitted to the ground by default in earth observation missions. 
Besides, the volume of these sensor data is negligible compared with images so we do not consider them when counting bpp.
We show more results of more metrics and baselines including other VAE-based methods in \appendixautorefname~\ref{add_res} which \sysname achieves the SOTA performance on all 6 perceptual metrics. 

\begin{figure}[htbp]
    \centering
    \includegraphics[width=0.99\linewidth]{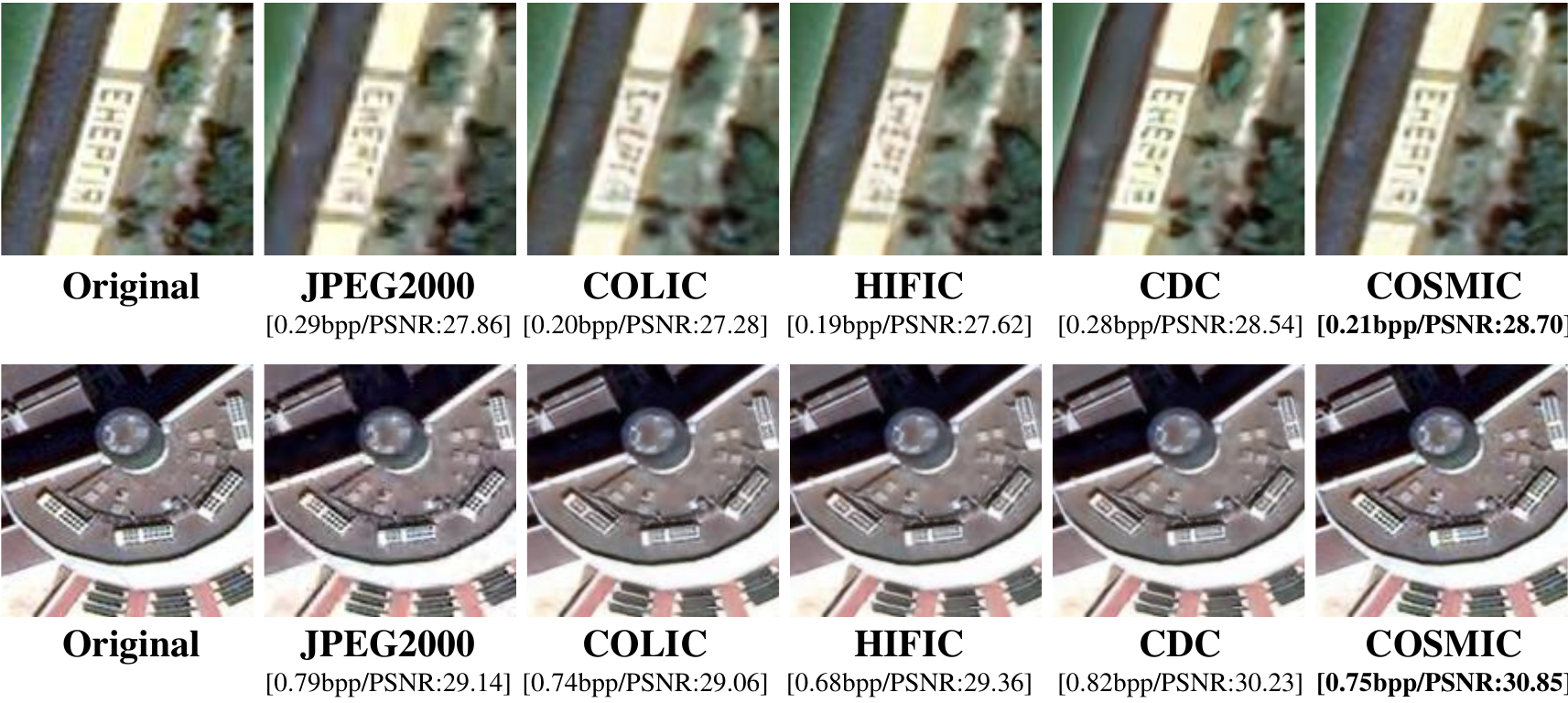}
    \vspace{-1ex}
    \caption{Decompressed fMoW images (full images in supplementary material). $1^{st}$ row: comparison under low bitrates, \sysname shows better visual effects. Compared with CDC, \sysname still gets slightly better visual reconstruction with less bitrates. $2^{nd}$ row: comparison under high bitrates. }
    \vspace{-2ex}
    \label{fig:visual_test1}
\end{figure}

\begin{figure}[!htbp]
    \centering
    \vspace{-2em}
    \includegraphics[width=0.98\linewidth]{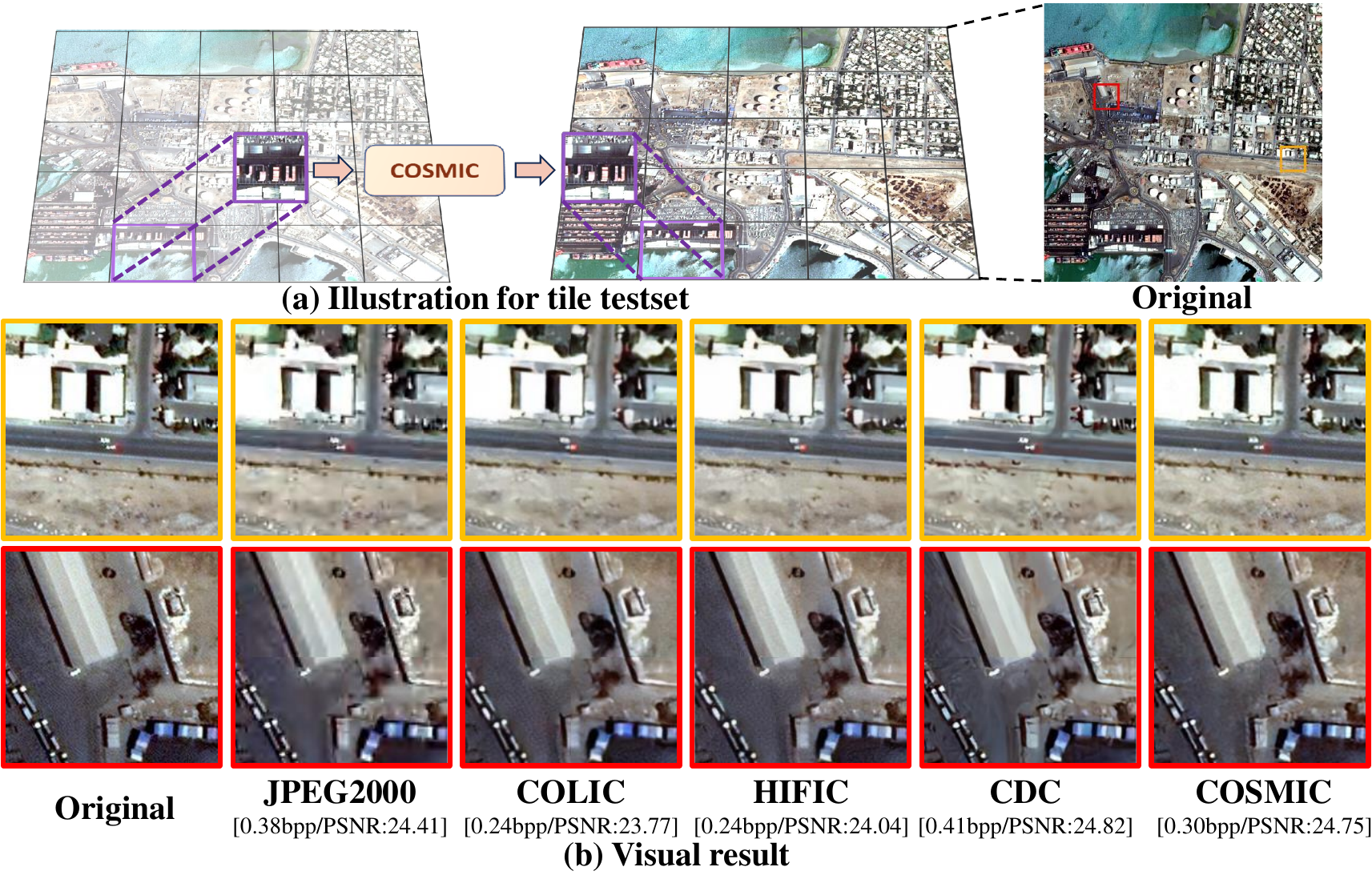}
    \caption{(a) Illustration of the tile test set. A high-resolution image is divided into many small sub-images (or patches), each of which is compressed individually. The reconstructed sub-images are then placed back in their original positions and stitched together to form a high-resolution reconstructed image. (b) On the tile test set, we provide two detailed views of a stitching area (outlined in \textcolor{orange}{orange} and \textcolor{red}{red}). The visual comparison between \sysname and the baseline shows that \sysname achieves the best visual effects in terms of texture alignment and consistency in color brightness.}
      \vspace{-2ex}
    \label{fig:visual_test2}
\end{figure}

\noindent\textbf{Visual results.} \figurename~\ref{fig:visual_test1} shows the example of reconstructed images at low bitrates and high bitrates.
For fair comparison, we only show the results of optimizing for image perception.
\figurename~\ref{fig:visual_test2} (b) shows an example of high-resolution image reconstruction by \sysname and baselines.
Due to the tiling and stitching process in image compression, we pay particular attention to the seams where different small patches form a larger image. 
JPEG2000 exhibits noticeable misalignment at the image seams. 
Hific and COLIC can not accurately restore the details of the seam. For example, in the picture outlined in orange, the car headlight at the seam has been reconstructed into a red dot.
The diffusion-based CDC reconstruction also exhibits some color differences between different sub-images and shows noticeable misalignment, such as the eaves at the seam in the picture outlined in red.
Compared to baselines, \sysname maintains a higher similarity in structure and color between different sub-images, resulting in significant visual improvements.

\subsection{Encoder efficiency analysis}

\begin{wraptable}{r}{0.65\textwidth}
\vspace{-2.5ex}
    \caption{Comparison of the on-satellite FLOPs with baselines on the tile test set. Best performances are highlighted in \textbf{bold}.} 
    \label{table:encoder_flops}
    \centering
    \resizebox{0.65\textwidth}{!}{
    \begin{tabular}{ccccccc}
\hline
Method  & FLOPs (G) & PSNR↑          & MS-SSIM↑       & LPIPS↓          & FID↓           & bpp↓          \\ \hline
CDC~\cite{yang2024lossy}     & 13.1             & 31.98          & 0.982 & 0.0462          & 45.49          & 0.66          \\
COLIC~\cite{li2022content}   & 26.4             & 29.04          & 0.975          & 0.0530          & 32.56          & 0.64          \\
Hific~\cite{mentzer2020high}   & 26.4             & 29.11          & 0.977          & 0.0384          & 27.01          & 0.60          \\
Elic~\cite{he2022elic}   &  21.78            &      \textbf{33.31}     &     \textbf{0.983}     &     0.0683     &      60.80     &      0.54     \\
HL\_RS~\cite{xiang2024remote}   &  11.87            &     31.30      &     0.979      &       0.0782    &    27.38       &     0.56      \\
\sysname & \textbf{4.9}     & 32.11 & 0.980    & \textbf{0.0359} & \textbf{13.50} & 0.59 \\ \hline
\end{tabular}
}
\end{wraptable}

We evaluate the lightweight encoder of \sysname in terms of FLOPs in \tablename~\ref{table:encoder_flops}. 
Compared with baselines, the \textit{on-satellite FLOPs} (including the encoder and entropy model) of \sysname has been significantly reduced by roughly $2.6\sim5 \times$ while the overall performance of \sysname can still outperform baselines under a similar bitrate.   

\subsection{Ablation study}
\label{sec:ablation}

\begin{figure}[!htbp]
    \centering
    \includegraphics[width=0.99\linewidth]{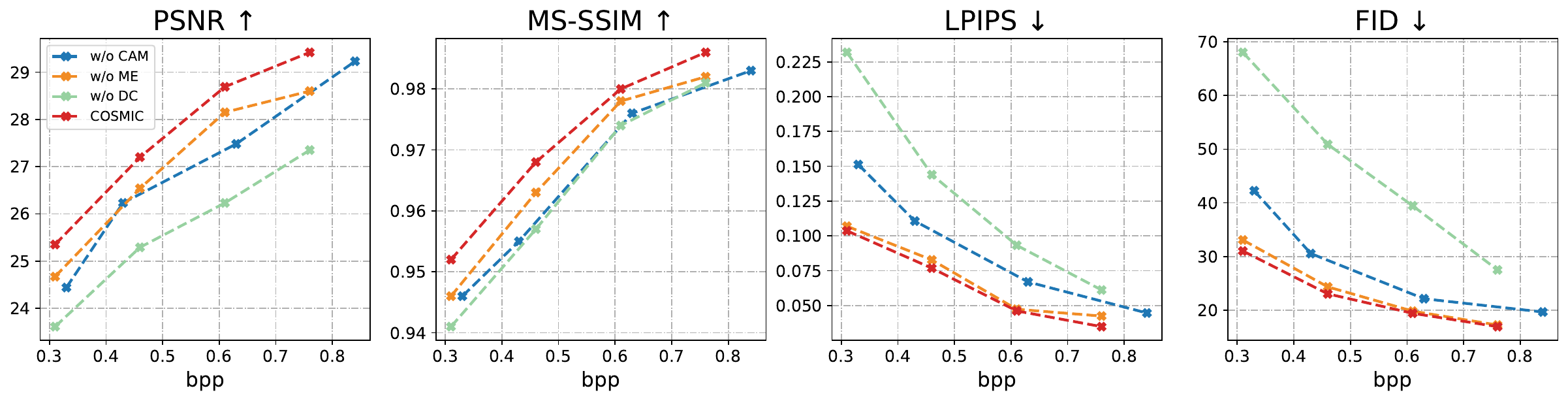}
    \caption{Ablation study with different variants of \sysname. ``w/o DC'' indicates that diffusion compensation is not used during decoding. ``w/o CAM''denotes the CAM module is removed in the encoder. ``w/o ME'' denotes that we remove the metadata encoder. }
    \label{fig:ablation}
\end{figure}

\begin{figure}[ht]
    \centering
    \includegraphics[width=0.99\linewidth]{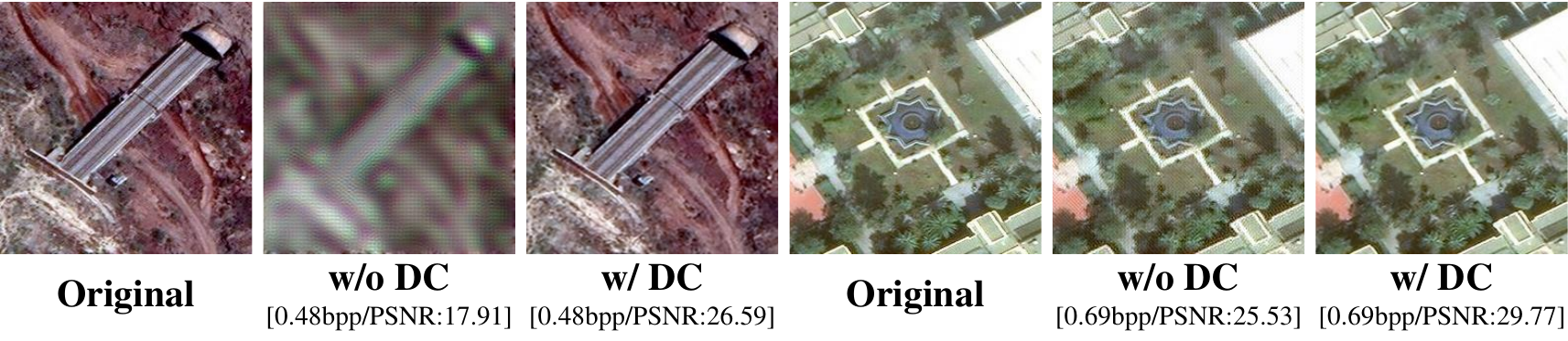}
    \caption{Contrast visual results. ``w/ DC'' means using of diffusion for compensation during decoding, while ``w/o DC'' indicates that diffusion compensation was not used during decoding.}
    \label{fig:visual_ablation}
\end{figure}

\noindent\textbf{w/o DC}.
To demonstrate the compensatory role of diffusion in the image reconstruction process, we remove the diffusion compensation module, and the results are shown in \figurename~\ref{fig:ablation}. 
We remove the compensation module and retrain the model to show the quantitative metrics.
The result shows that the diffusion model plays an important role in decompressing images to get better perceptual metrics.
To show the compensatory role of the diffusion model more clearly, we directly remove the diffusion model and only use the output information of the lightweight encoder $\mathcal{E}$ to reconstruct the image.
We find that at low bitrate, diffusion compensation is more important. 
Due to the insufficient feature extraction capability of the lightweight encoder, many image content details are lost to save the transmission rate, and diffusion needs to reconstruct much of the image content guided by the limited output of the encoder, along with the metadata. 
As the bitrate increases, more features extracted by the encoder can be retained, and at this point, diffusion only needs to compensate for some image details that the encoder failed to capture. 
The visual results are shown in \figurename~\ref{fig:visual_ablation}.
The results indicate that using the diffusion model as compensation is very useful, especially with a small bitrate. 

\noindent\textbf{w/o CAM}.
To show the CAM module can capture non-local information, which can help the encoder $\mathcal{E}$ to get higher quality representation. 
We remove the CAM module and show the results in \figurename~\ref{fig:ablation}. 
The result shows that CAM module is effective and can achieve better RD performance.

\noindent\textbf{w/o ME}.
To show that sensor data can guide the generation of diffusion, we remove the metadata encoder and only use image encoding for diffusion generation. 
Please note that there are mainly two kinds of sensor data here including the camera parameters such as the off-nadir angle and target azimuth used during image capture and the data on cloud cover, illumination, and ground sample distance, etc. 
\figurename~\ref{fig:ablation} indicates that sensor data helps guide the diffusion in reconstructing the image.

\begin{figure}[!htbp]
    \centering
    
    \includegraphics[width=0.95\linewidth]{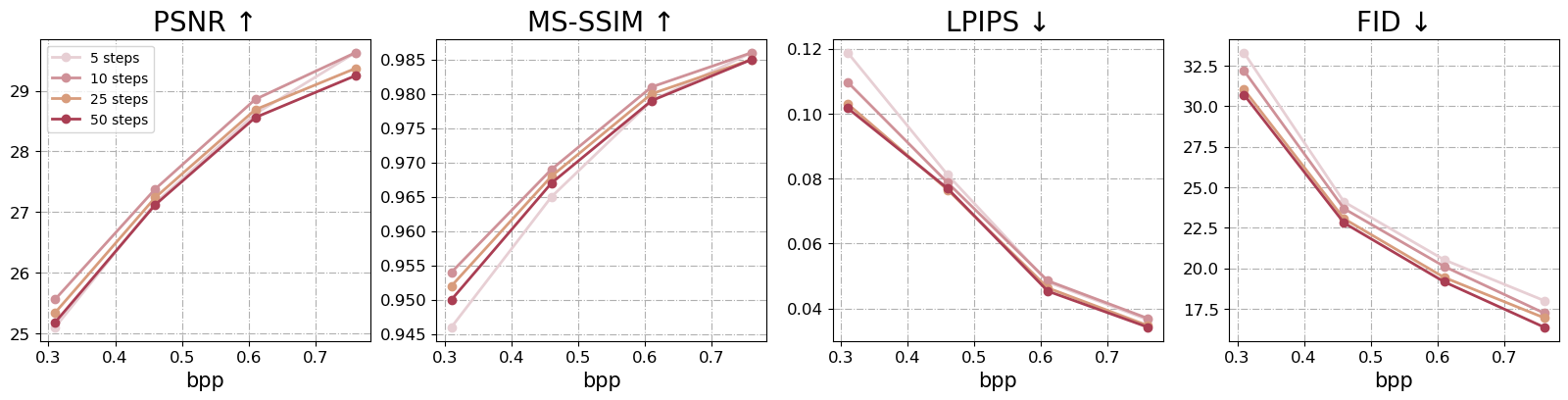}
    \caption{Compression performance with different numbers of decoding step.}
    \vspace{-1ex}
    \label{fig:step_ablation}
\end{figure}

\noindent\textbf{Influence of decoding steps}.
We further investigate the impact of different denoising step counts in the reverse diffusion process on the reconstructed image quality, as shown in \figurename~\ref{fig:step_ablation}. 
We find that as the number of denoising steps increases, the perceptual metrics of the generated images gradually improve, which is consistent with the exploration of the relationship between denoising steps and FID scores in DDIM~\cite{song2020denoising}. 
While the perceptual results improve, distortion metrics such as PSNR experience a slight decline, showcasing the trade-off between perceptual quality and distortion. 
In practical deployment and downstream tasks, the number of denoising steps can be selected based on different emphases on distortion and perceptual performance to suit actual applications. 

\begin{table}[t]
\vspace{-1em}
\centering
\caption{Effect on image classification model.}
    \label{table:case_study}
    \centering
    \resizebox{0.95\textwidth}{!}{
    \begin{tabular}{ccccccccc}
    \hline
        Classes & Original & JPEG2K~\cite{taubman2002jpeg2000} & Elic~\cite{he2022elic} & COLIC~\cite{li2022content} & HIFIC~\cite{mentzer2020high} & CDC~\cite{yang2024lossy} & HL\_RS~\cite{xiang2024remote} & \sysname \\ \hline
        10 & 98.95\% & -4.21\% & -5.27\% & -1.06\% & -1.06\% & -2.11\% &-2.11\% & -1.06\% \\ 
        15 & 97.92\% & -4.17\% & -3.84\% & -0.70\% & -0.70\% & -1.40\% & -1.39\% & -0.70\% \\ 
        20 & 98.42\% & -3.16\% & -3.68\% &-0.53\% & -0.53\% & -1.06\% & -2.1\% &-1.06\% \\ \hline
    \end{tabular}
    }
\end{table}

\subsection{Compression influence on downstream tasks}

To confirm that \sysname will not affect the downstream remote sensing tasks, such as image classification, we choose the image classification task in~\cite{8844264} to show the effect caused by compression, as shown in Table~\ref{table:case_study}.
The same test images are compressed at a low bitrate level and decompressed, then processed through the classification model to obtain the accuracy changes on different numbers of classes. The result shows that JPEG2000 and Elic reduce the accuracy by the most and \sysname is outstanding in learning methods.

\section{Conclusion}
\label{sec:conclusion}

We present \sysname, a novel approach to compress images for satellite earth observation missions. 
We first design a lightweight encoder to adapt to the limited resources on satellites. 
Then, we introduce a conditional latent diffusion model by using the sensor data of satellite as instructions to compensate the missing details due to the lightweight encoder's degradation of feature extraction. 
Extensive results indicate that \sysname can not only achieve SOTA compression performance for 2 typical satellite tasks but also guarantee the accuracy of satellite images' downstream tasks.

\noindent\textbf{Limitations \& Future work.} Even though \sysname compensates for encoder limitations using diffusion, at extremely low bpp (e.g., less than 0.1bpp), the information provided by latent image coding may not be enough to support diffusion in generating high-fidelity images. 
We think the main reason is that we only finetune the pretrained Stable Diffusion model, which lacks sufficient prior knowledge of satellite images. As a prospective solution, we will train a diffusion model specifically for satellite images or use historical satellite images as a reference for further improvements.

\noindent\textbf{Acknowledgement.} This work was supported by the National Key R\&D Program of China (2022YFB3105202), National Natural Science Foundation of China (62106127, 62301189, 62132009), and key fund of National Natural Science Foundation of China (62272266).

\clearpage
\bibliographystyle{plainnat}
\bibliography{ref}

\clearpage
\appendix

\section{Details about \sysname}
\label{appendix:detail}

\subsection{Details of train and inference}

We present a more detailed explanation of our two-stage training and inference pipeline in \algorithmautorefname~\ref{algo:train} and \algorithmautorefname~\ref{algo:inference}.

\begin{algorithm}[h]
\SetAlgoLined
 
 \SetKwInOut{Input}{input}\SetKwInOut{Output}{output}
 \SetKwInput{KwParam}{Parameters}
 \SetKwRepeat{KwRepeat}{Repeat}
 
 \Input{A satellite image $\mathbf{x}$, referenced image $\mathbf{x}$(the same as satellite image, used only for training), metadata $\mathrm{m}_{\mathrm{i}}\in \mathbb{R} ^{\mathrm{M}}$}
 \Output{Reconstructed image $\hat{\mathbf{x}}$}
 \KwParam{Image encoder on satellite $\mathcal{E}$, image encoder of diffusion $\tilde{\mathcal{E}}$, image decoder $\mathcal{D}$, noise prediction network $\mathrm{\epsilon}_{\mathrm{\theta}}$}
 \BlankLine
\tcc{\emph{Training stage 1}: Train $\mathcal{E}$, $\tilde{\mathcal{E}}$, and $\mathcal{D}$ together.}
$\mathbf{y} = \mathcal{E} \left( \mathbf{x} \right) $\;
$\mathrm{\zeta}=\mathrm{entropy}\_model(\mathbf{y})$\;
$\mathrm{\mu}_{\mathrm{z}},\mathrm{\sigma}_{\mathrm{z}}\,\,=\,\,\tilde{\mathcal{E}}\left( \mathrm{x}_0 \right) 
$\;
$\mathrm{\varepsilon}_{\mathrm{z}}\sim \mathcal{N} \left( 0,\mathrm{I} \right)$\;
$\mathrm{z}_0=\mathrm{\mu}_{\mathrm{z}}+\mathrm{\sigma}_{\mathrm{z}}*\mathrm{\varepsilon}_{\mathrm{z}}$\;
$\hat{\mathbf{x}}=\mathcal{D} \left( \mathrm{concat}\left( \mathrm{z}_0, \mathrm{deconv}\left( \lfloor \mathbf{y} \rceil  \right) \right) \right) $\;
optimize the parameters of $\mathcal{E}$, $\tilde{\mathcal{E}}$, and $\mathcal{D}$ following: $\mathcal{L} _{\mathrm{IC}}=\mathrm{R}+\mathrm{\lambda D}=\mathbb{E} \left[ -\log _2\mathrm{p}\left( \lfloor \mathbf{y} \rceil|\mathrm{\zeta} \right) -\log _2\mathrm{p}\left( \mathrm{\zeta} \right) \right] +\mathrm{\lambda}\mathbb{E} \left[ \mathrm{d}\left( \mathbf{x},\hat{\mathbf{x}} \right) \right]$

\BlankLine
\tcc{\emph{Training stage 2}:Freeze the parameters of $\mathcal{E}$, $\tilde{\mathcal{E}}$, and $\mathcal{D}$, and only update the parameters of $\mathrm{\epsilon}_{\mathrm{\theta}}$.}
\Repeat{converge}{
sample $\mathrm{x}, \mathrm{m}\sim \mathrm{dataset}$\;
$\mathrm{t}\sim \mathcal{U} \left( 1,2,...,\mathrm{T} \right)$\;
$\mathrm{\epsilon}_{\mathrm{t}}\sim \mathcal{N} \left( 0, \mathrm{I} \right)$\;
$\mathrm{c}^{\mathrm{final}}=\mathrm{metadata}\_encoder(m)$\;
optimize the parameters of $\mathrm{\epsilon}_{\mathrm{\theta}}$ following:
$\mathcal{L} _{\mathrm{ldm}}=\mathbb{E} _{\mathrm{t},\mathrm{z}_0,\mathrm{\epsilon}\sim \mathcal{N} \left( 0,1 \right)}\left[ \left\| \bm{\epsilon}_{\mathrm{t}}-\bm{\epsilon}_{\mathrm{\theta}}\left( \mathrm{z}_{\mathrm{t}},\mathrm{t},\lfloor \mathbf{y} \rceil ,\mathrm{c}^{\mathrm{final}} \right) \right\| ^2 \right] $
}

 \caption{Two-stage training of \sysname.}\label{algo:train}
\end{algorithm}

\begin{algorithm}[h]
\SetAlgoLined
\SetKwInOut{Input}{input}\SetKwInOut{Output}{output}
\SetKwInput{KwParam}{Parameters}
\SetKwRepeat{KwRepeat}{Repeat}

\Input{A satellite image $\mathbf{x}$, metadata $\mathrm{m}_{\mathrm{i}}\in \mathbb{R} ^{\mathrm{M}}$}
\Output{Reconstructed image $\hat{\mathbf{x}}$}
\KwParam{Image encoder on satellite $\mathcal{E}$, image decoder $\mathcal{D}$, noise prediction network $\mathrm{\epsilon}_{\mathrm{\theta}}$}
\BlankLine

\tcc{\emph{Compress}}
$\mathbf{y} = \mathcal{E} \left( \mathbf{x} \right) $\;
$\mathrm{\zeta}=\mathrm{entropy}\_model(\mathbf{y})$\;
$\lfloor \mathbf{y} \rceil =\,\,\mathrm{Quantization}\left( \mathrm{y},\mathrm{\zeta} \right)$\;

\tcc{\emph{Decompress}}
sample $\mathrm{z}_{\mathrm{T}}\sim \mathcal{N} \left( 0,\mathrm{I} \right)$\;
\For{$\mathrm{t}=\mathrm{T},...,1
$}{
$\mathrm{\epsilon}_{\mathrm{t}} \leftarrow \mathrm{\epsilon}_{\mathrm{\theta}}\left( \mathrm{z}_{\mathrm{t}},\mathrm{t},\lfloor \mathbf{y} \rceil,\mathrm{c}^{\mathrm{final}} \right) 
$\;
$\mathrm{z}_{\mathrm{t}-1} \leftarrow \mathrm{DDIM}\left( \mathrm{z}_{\mathrm{t}},\mathrm{t},\mathrm{\epsilon}_{\mathrm{t}} \right) 
$\;
}
return $\mathrm{z}_0\prime$\;
$\hat{\mathbf{x}}=\mathcal{D} \left( \mathrm{concat}\left( \mathrm{z}_0\prime, \mathrm{deconv}\left( \lfloor \mathbf{y} \rceil  \right) \right) \right) $\;

\caption{Compress and decompress pipeline of \sysname.}\label{algo:inference}
\end{algorithm}

\section{Additional Rate-Distortion(Perception) Results}
\label{add_res}

\begin{figure}[!ht]
    \centering
    \includegraphics[width=0.99\linewidth]{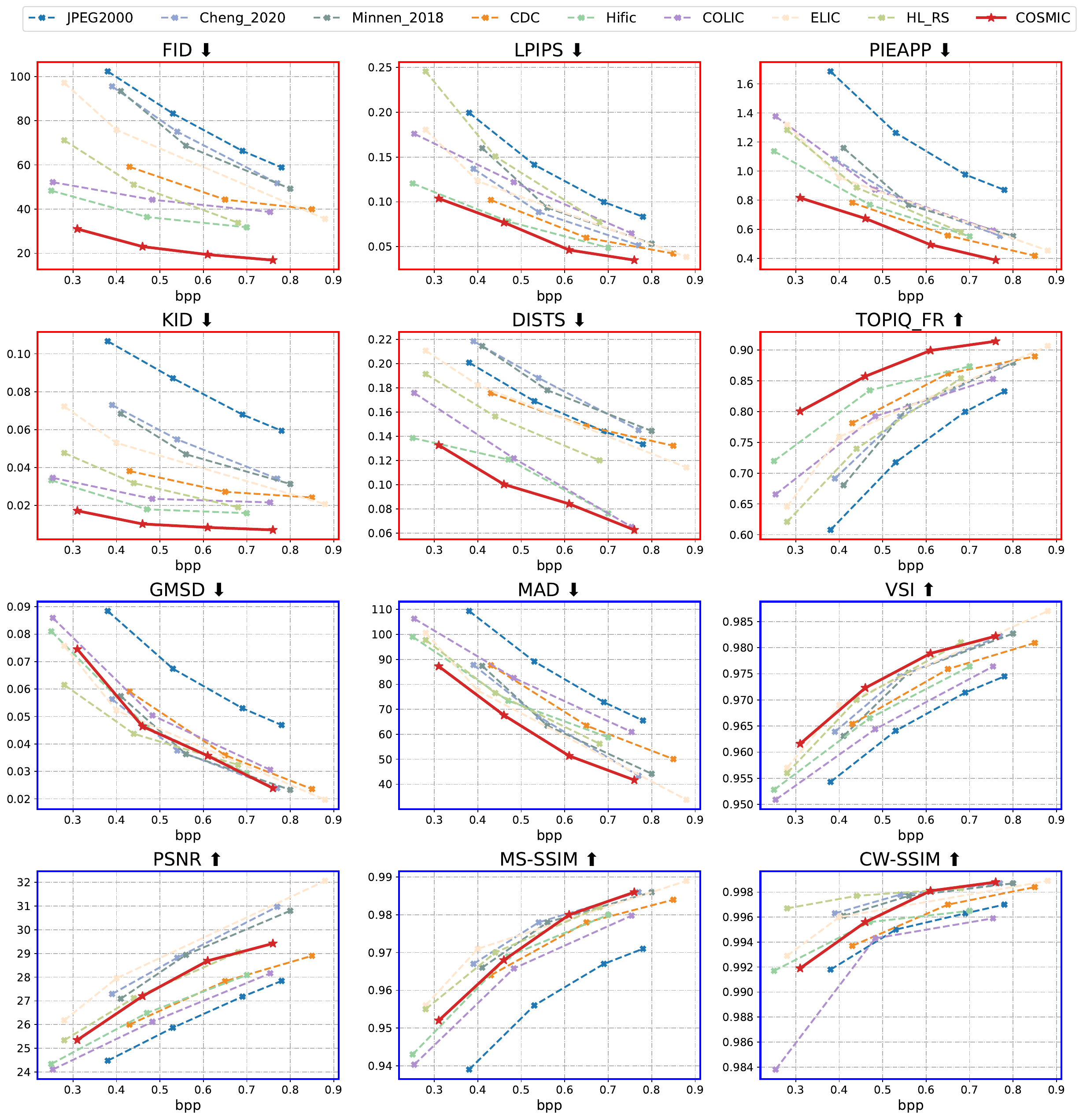}
    \caption{Rate-Distortion(Perception) for fMoW dataset.}
    \label{fig:more_RD}
    \vspace{-3ex}
\end{figure}

\section{Failure cases analysis}

\begin{figure}[!ht]
    \centering
    \includegraphics[width=0.99\linewidth]{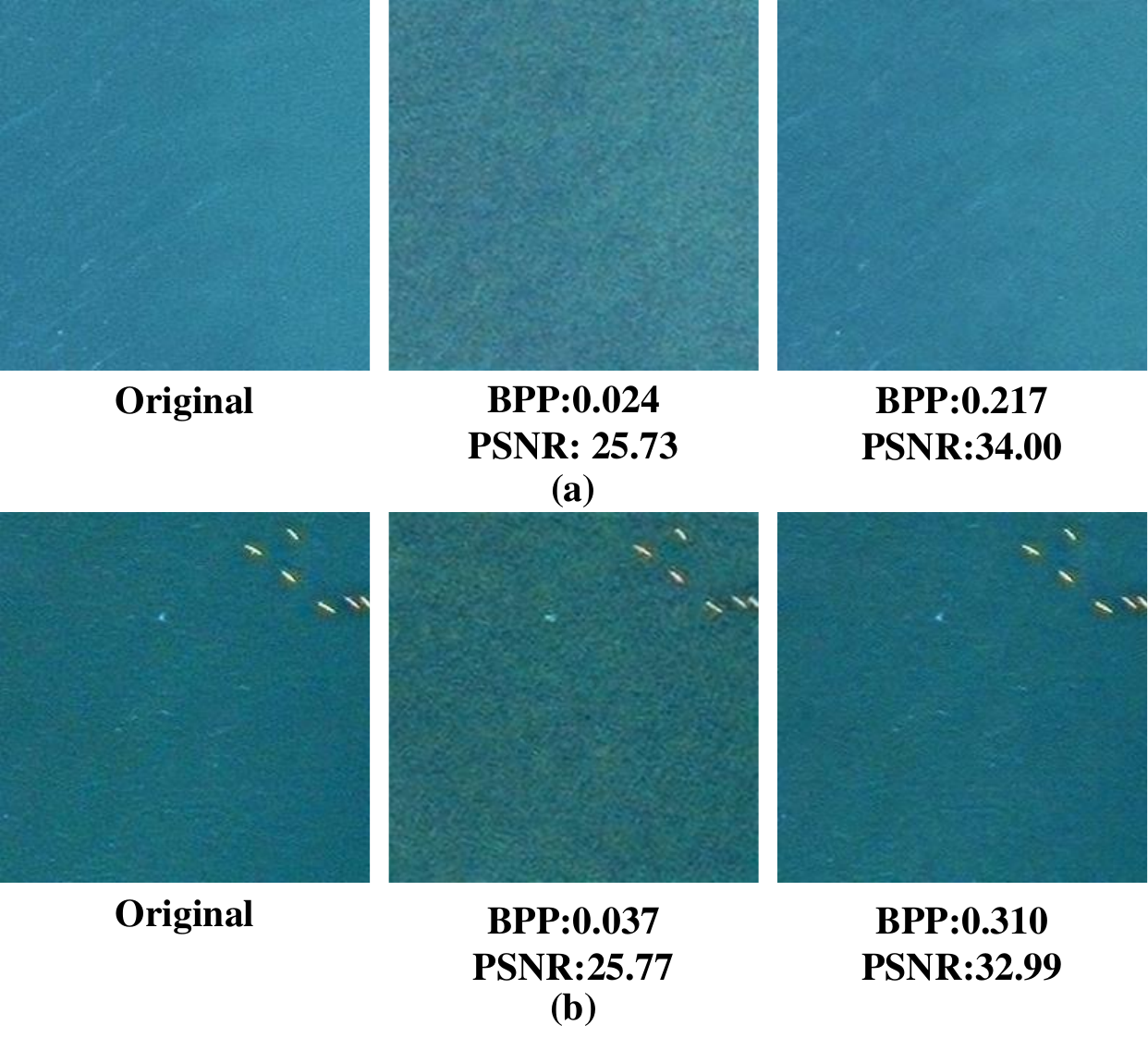}
    \caption{Failure examples.}
    \label{fig:failure}
    \vspace{-3ex}
\end{figure}

Stable diffusion was originally used for generative tasks, which can produce textures and content that do not exist in reality.
In \sysname, we ensure the consistency of content and texture by injecting discrete latent coding $\lfloor \mathbf{y} \rceil$ into the Vanilla Convolution (VC) blocks of the noise prediction network to provide structural information.
However, in some cases, especially at extremely low bitrates, the latent discrete coding may lack structural information, leading to the diffusion generating non-existent textures. 
As illustrated in \figureautorefname~\ref{fig:failure} (a), when a satellite image captures an ocean scene, the insufficient information in the latent coding at low bitrates requires heavy reliance on diffusion for generation, resulting in textures that do not exist in the reconstructed image. 
In the case of high bitrates, as described in the main paper, the reliance on diffusion reduces, and in the same scenario, the appearance of non-existent textures can be controlled.
However, in practice, remote sensing downstream tasks are more concerned with regions of interest (ROIs). 
For example, in remote sensing tasks for oceans, such as ship detection
, only the parts where ships are present are of interest. 
As shown in \figureautorefname~\ref{fig:failure} (b), even at low bitrates, while the reconstruction of the sea surface may introduce non-existent textures, the reconstruction of the ship parts remains intact. 
Therefore, although \sysname may have minor visual flaws, it does not affect downstream detection tasks.

\section{Why not artifact correction for JPEG2000?}
\begin{figure}[!ht]
    \centering
    \includegraphics[width=0.99\linewidth]{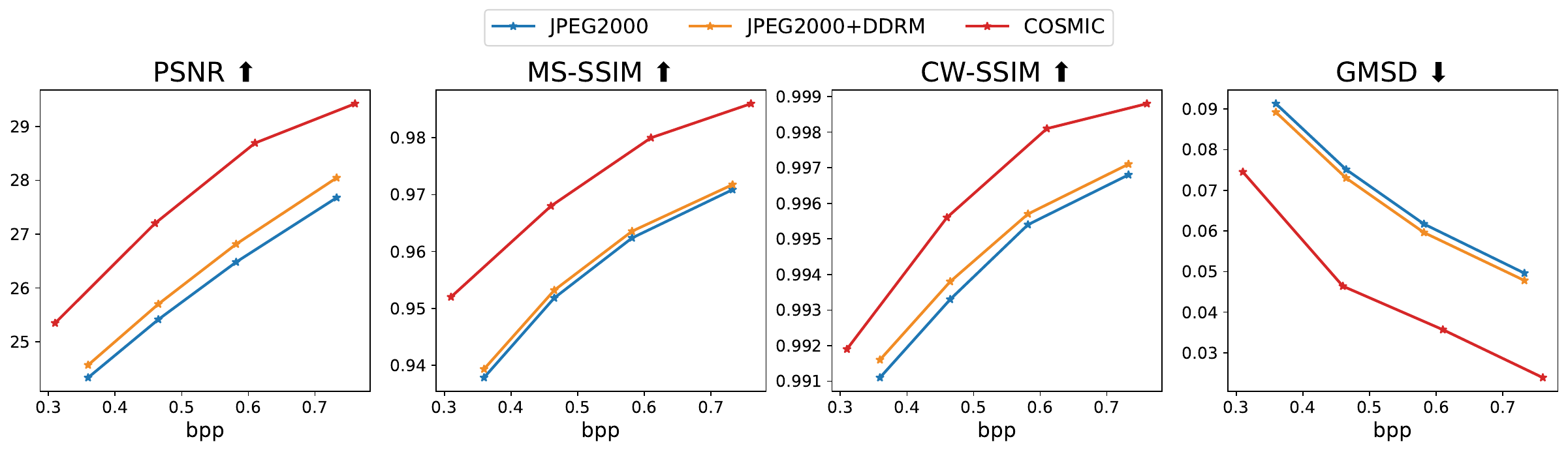}
    \caption{Comparison results of \sysname, JPEG2000 and JPEG2000+DDRM.}
    \label{fig:jpeg_enh}
\end{figure}

The most direct way to improve the quality of satellite images is to use artifact correction on JPEG2000 compressed images.
This approach does not introduce the computational burden on the satellite and can also improve image quality to a certain extent.
We use DDRM~\cite{kawar2022denoising}, which is an image restoration method, for JPEG2000 compressed images to remove the artifact.
As shown in \figurename~\ref{fig:jpeg_enh}, JPEG2000+DDRM is slightly better than JPEG2000 but the improvement is limited, and there is still a big gap with \sysname.
We believe that this is mainly due to the fact that traditional compression methods based on static compression procedures cannot fit the data well like neural networks, and therefore lose a large amount of information, making image restoration methods difficult to improve the image quality.
As a result, it is necessary to design a learned compression method to improve satellite image compression quality without introducing excessive computational burden on the satellite.

\section{Influence of random seed}

To show the influence of different initial gaussian noise on the quality of reconstructed images, we randomly select 5 different seeds for each bit rate and show the 2-Sigma Error Bars results in \tableautorefname~\ref{table:random_seed}.
The results demonstrate that our method has small variances in various quantitative metrics, proving the robustness of our method to initialized random Gaussian noise.

\begin{table}
    \caption{The results of different random seeds for fMoW dataset.}
    \label{table:random_seed}
    \centering
    \resizebox{0.9\textwidth}{!}{
    \begin{tabular}{c|cccc}
\hline
bpp  & PSNR & MS-SSIM & LPIPS & FID \\ \hline
0.31 &  25.3698 $\pm$ 0.0447    &  0.9521 $\pm$ 0.0004      & 0.1040 $\pm$ 0.0010      &  31.2750 $\pm$ 0.4328   \\
0.46 &  27.2411 $\pm$ 0.0222   &  0.9682 $\pm$ 0.0004      & 0.0761 $\pm$ 0.0006     & 23.0243 $\pm$ 0.2253   \\
0.61 &  28.6654 $\pm$ 0.0251    &  0.9799 $\pm$ 0.0004       &   0.0461 $\pm$ 0.0004   &  19.4264 $\pm$ 0.1059  \\
0.76 &  29.3617 $\pm$ 0.0608   &  0.9852 $\pm$ 0.0006      &  0.0363 $\pm$ 0.0007    &  16.9090 $\pm$ 0.1683  \\ \hline
\end{tabular}
    }
\end{table}

\section{Additional visualization of reconstructed images}

\begin{figure}
    \centering
    \includegraphics[width=0.95\linewidth]{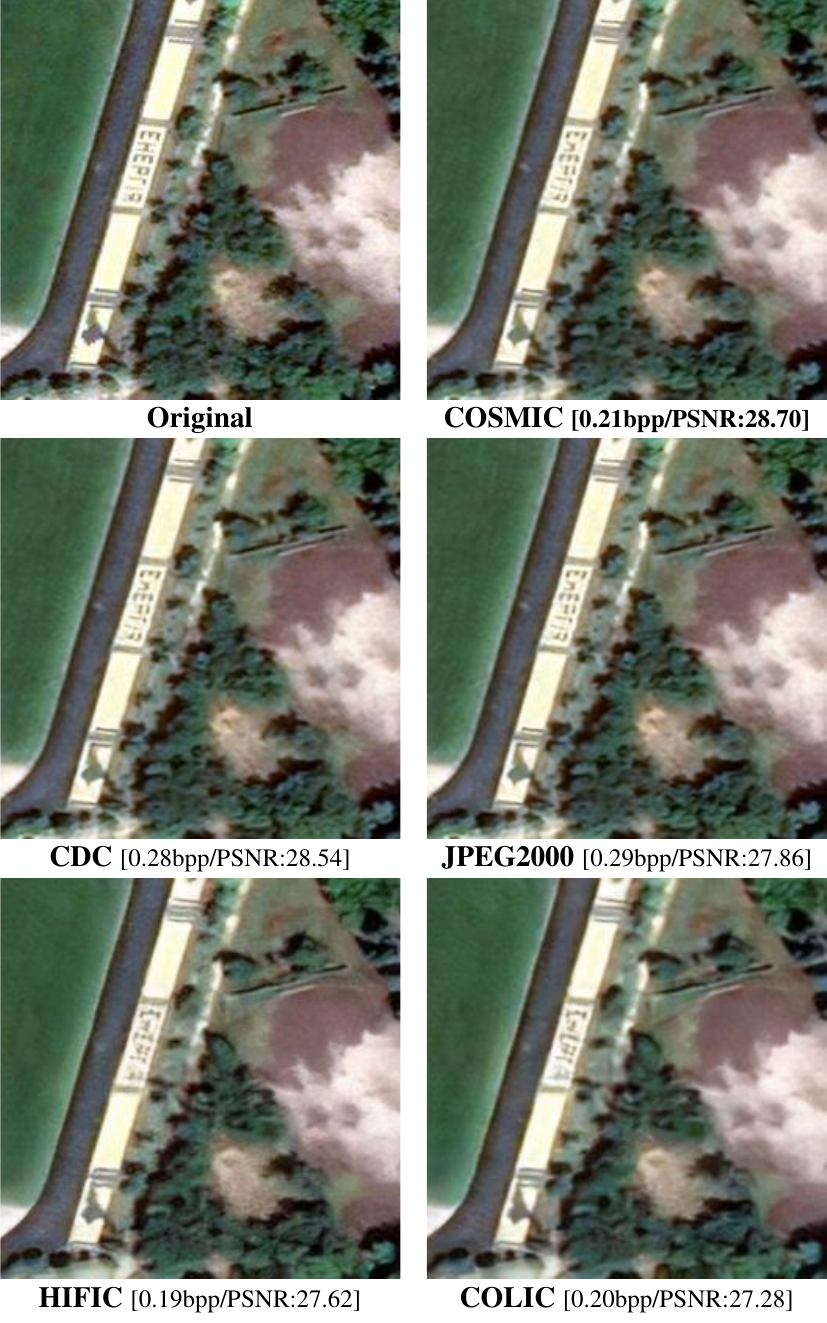}
    \caption{Reconstructed fMoW images under low bitrates.}
    \label{fig:show_low}
\end{figure}

\begin{figure}[!ht]
    \centering
    \includegraphics[width=0.7\linewidth]{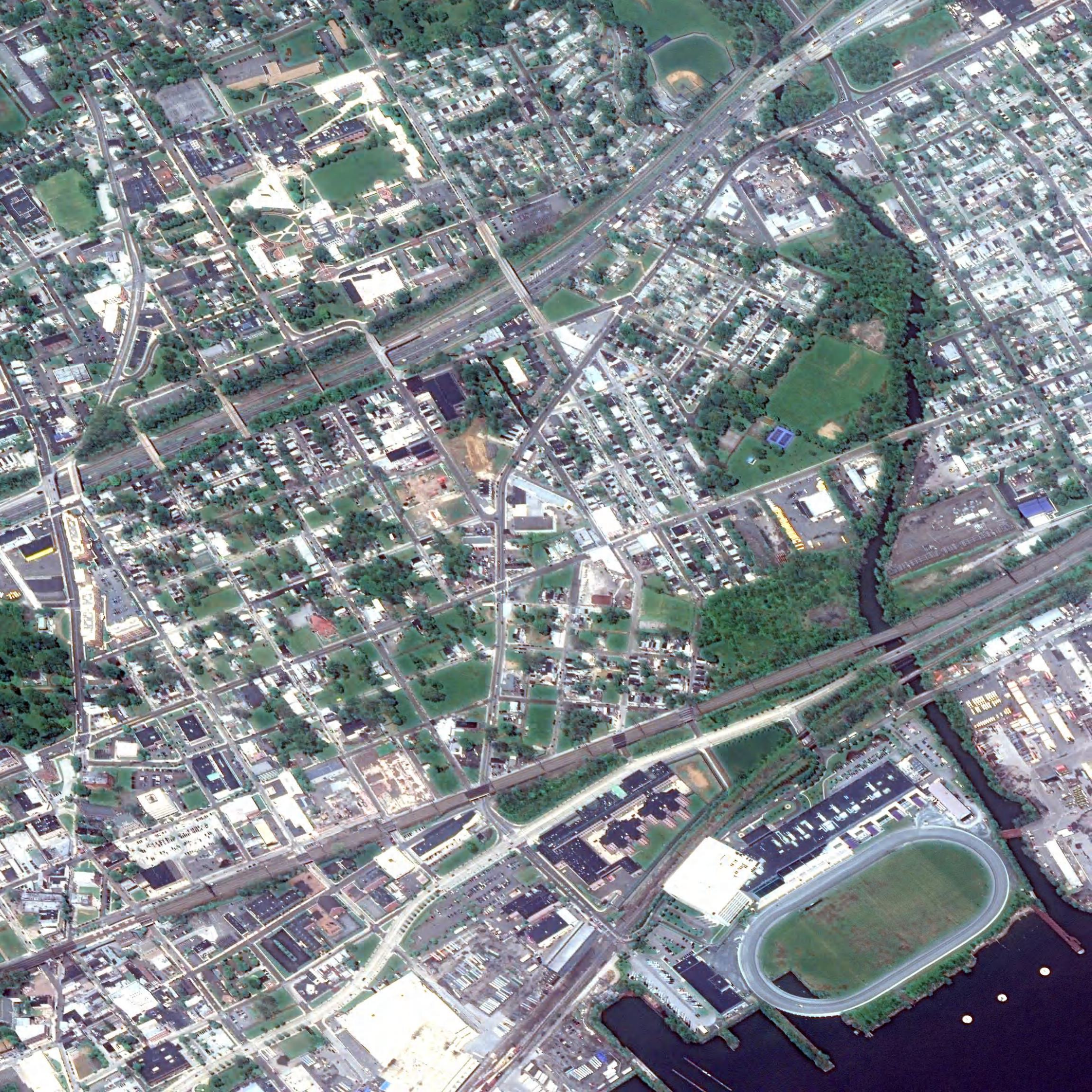}
    \caption{\textbf{High bpp}: JPEG2000, PSNR=28.95, bpp=0.69.}
    \label{fig:jpeg20002}
\end{figure}

We show the full images of \figureautorefname~\ref{fig:visual_test1} and \figureautorefname~\ref{fig:visual_test2}in the main paper, and give more visualization results, as shown in \figureautorefname~\ref{fig:show_low} $\sim$ ~\ref{fig:jpeg20002}.
We select visualization results for two different test sets under higher bitrates and lower bitrates from each dataset. 
For the tile test set, under lower bitrates, as shown in \figureautorefname~\ref{fig:visual_test2} in the main paper, the reconstructed images by the baselines exhibit noticeable discontinuities at the stitching seams. 

\begin{figure}
    \centering
    \includegraphics[width=0.95\linewidth]{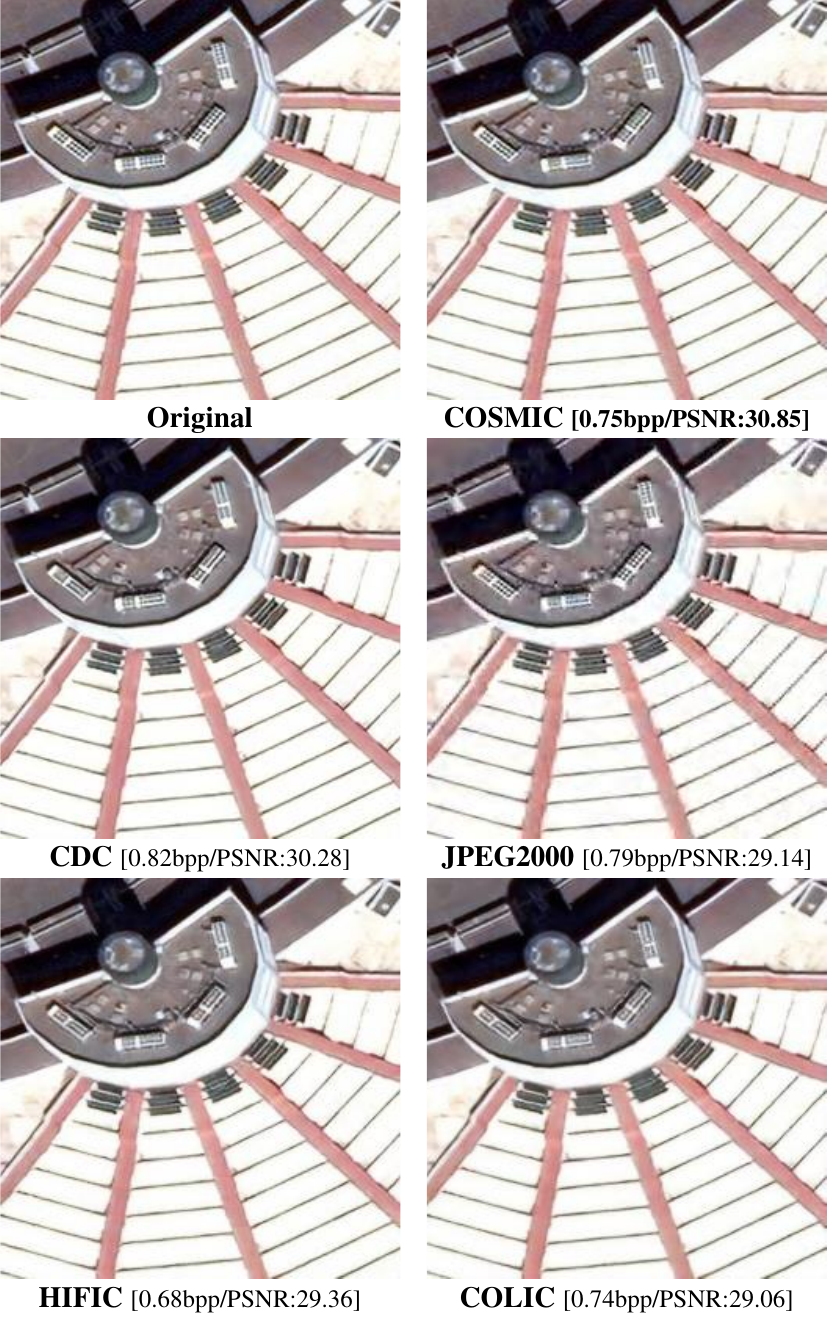}
    \caption{Reconstructed fMoW images under high bitrates.}
    \label{fig:show_hi}
\end{figure}

\begin{figure}[!ht]
    \centering
    \includegraphics[width=0.7\linewidth]{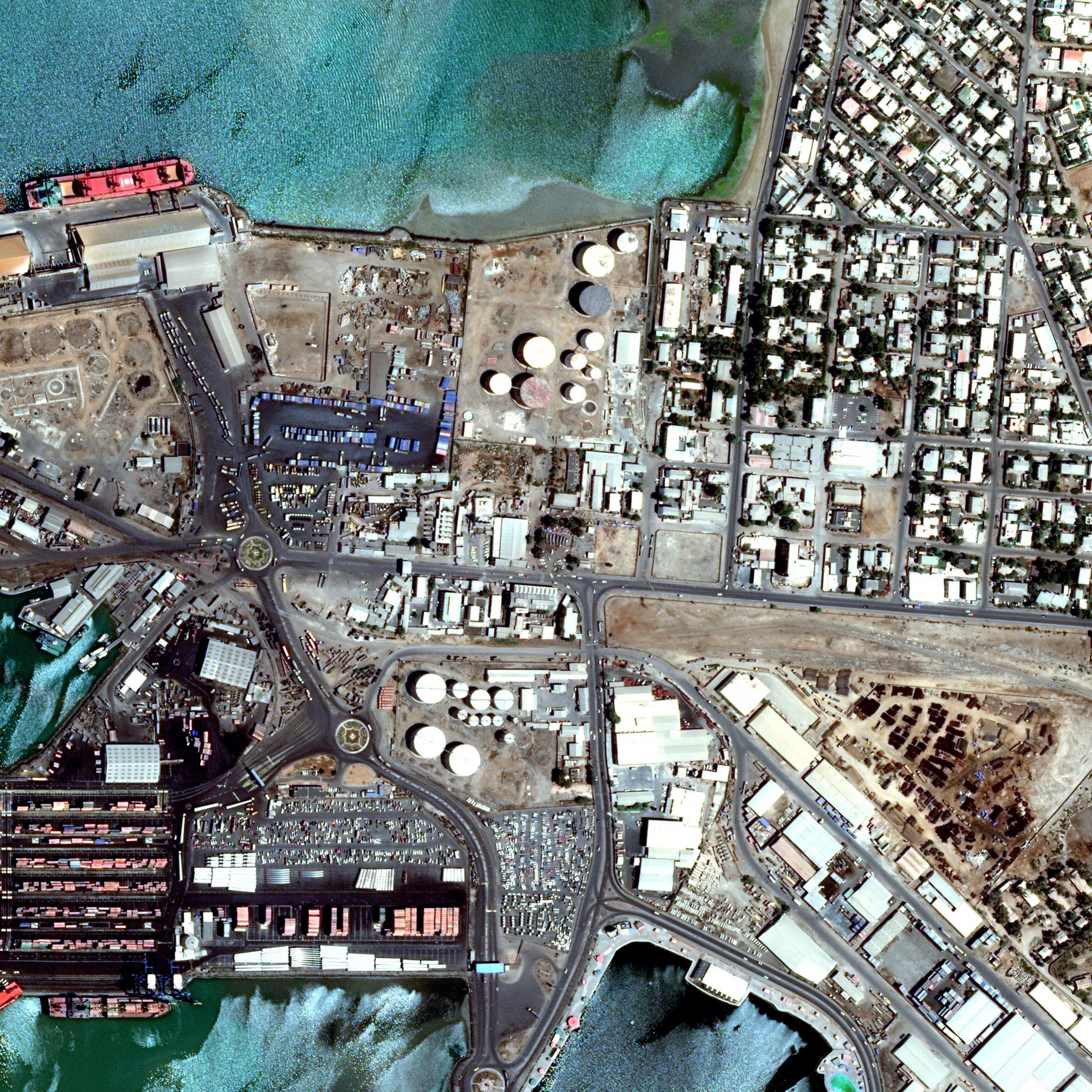}
    \caption{Ground truth.}
    \label{fig:gt}
\end{figure}

\begin{figure}[!ht]
    \centering
    \includegraphics[width=0.7\linewidth]{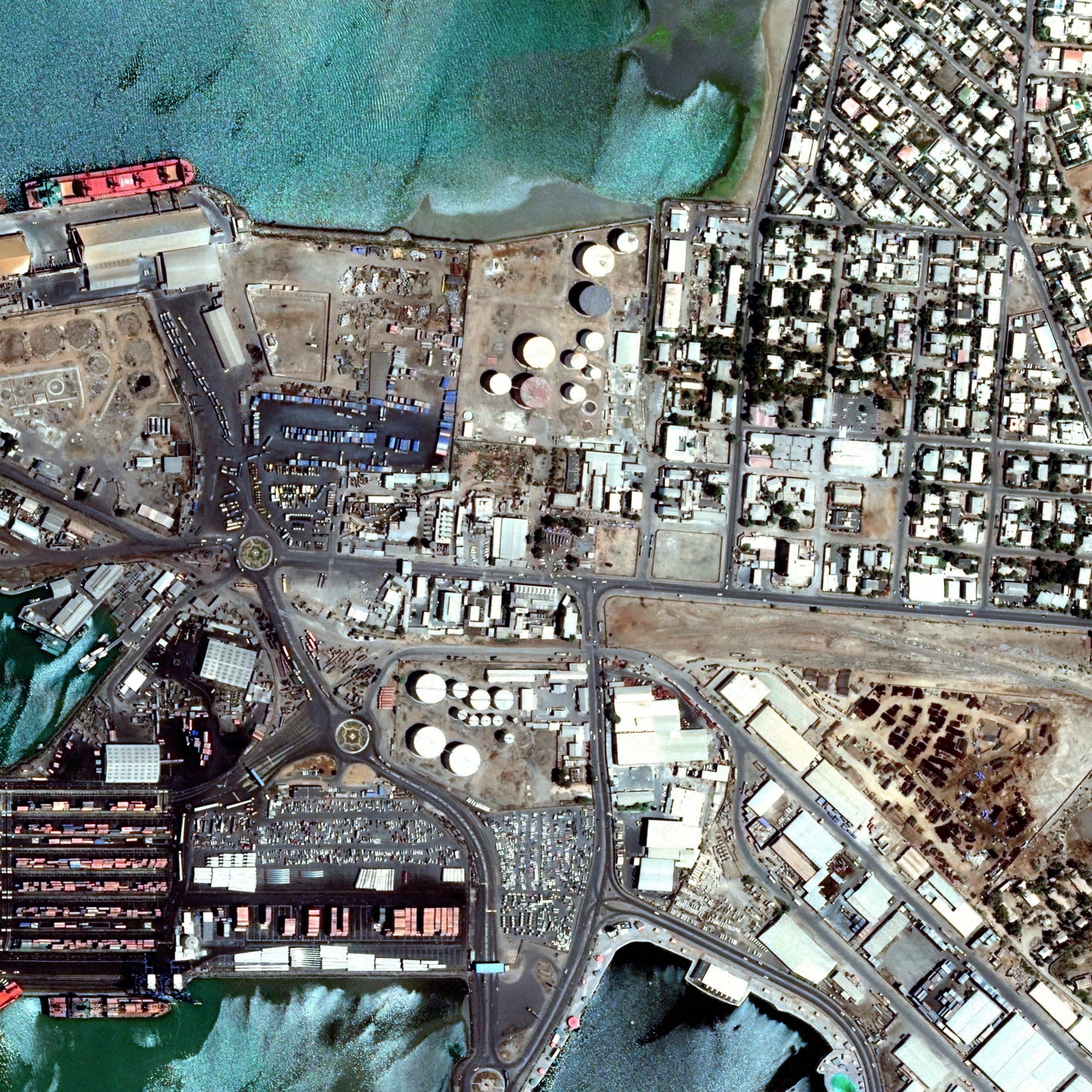}
    \caption{\textbf{Low bpp}: \sysname, PSNR=24.75, bpp=0.30.}
    \label{fig:our}
\end{figure}

\begin{figure}[!ht]
    \centering
    \includegraphics[width=0.7\linewidth]{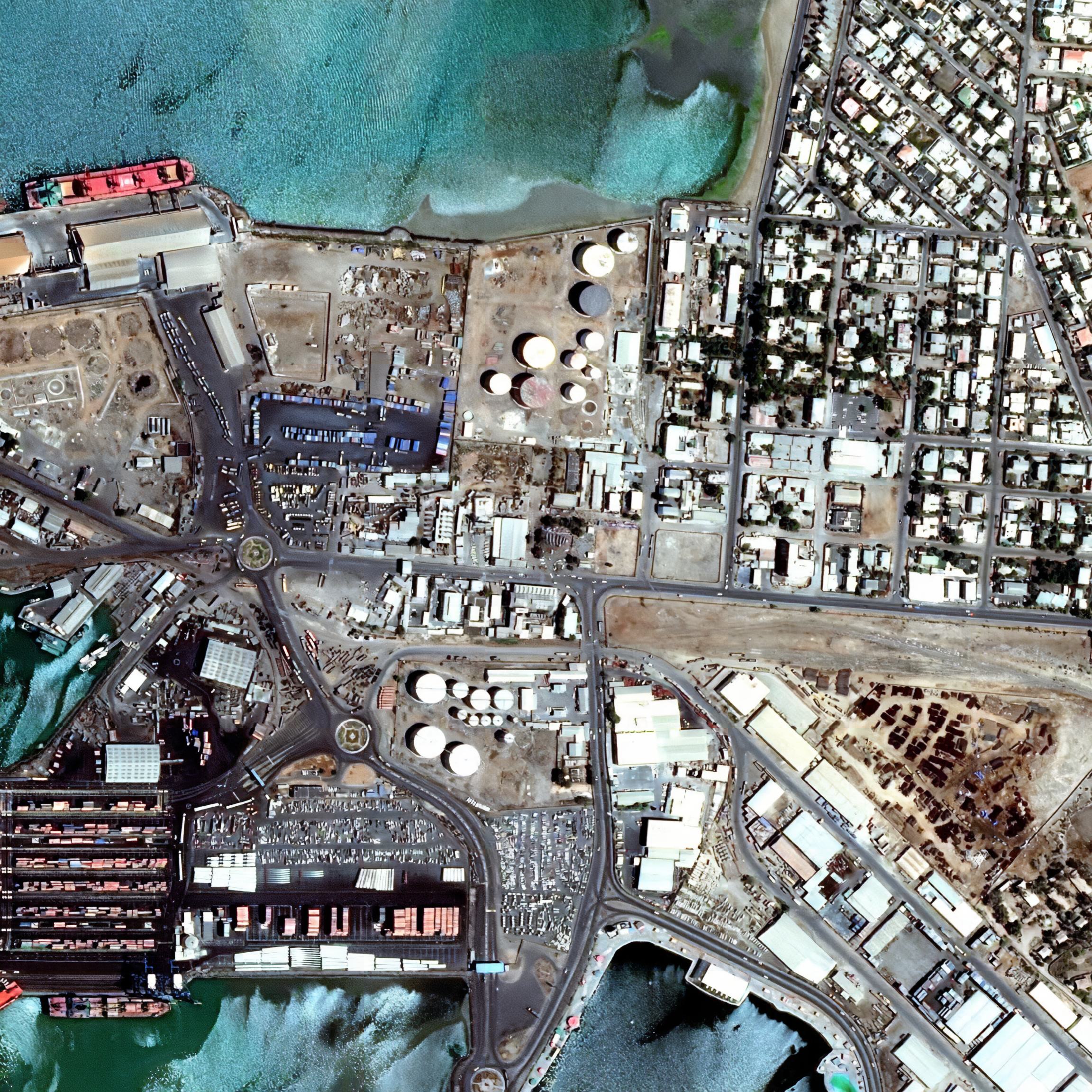}
    \caption{\textbf{Low bpp}: HIFIC, PSNR=24.04, bpp=0.24.}
    \label{fig:hific}
\end{figure}

\begin{figure}[!ht]
    \centering
    \includegraphics[width=0.7\linewidth]{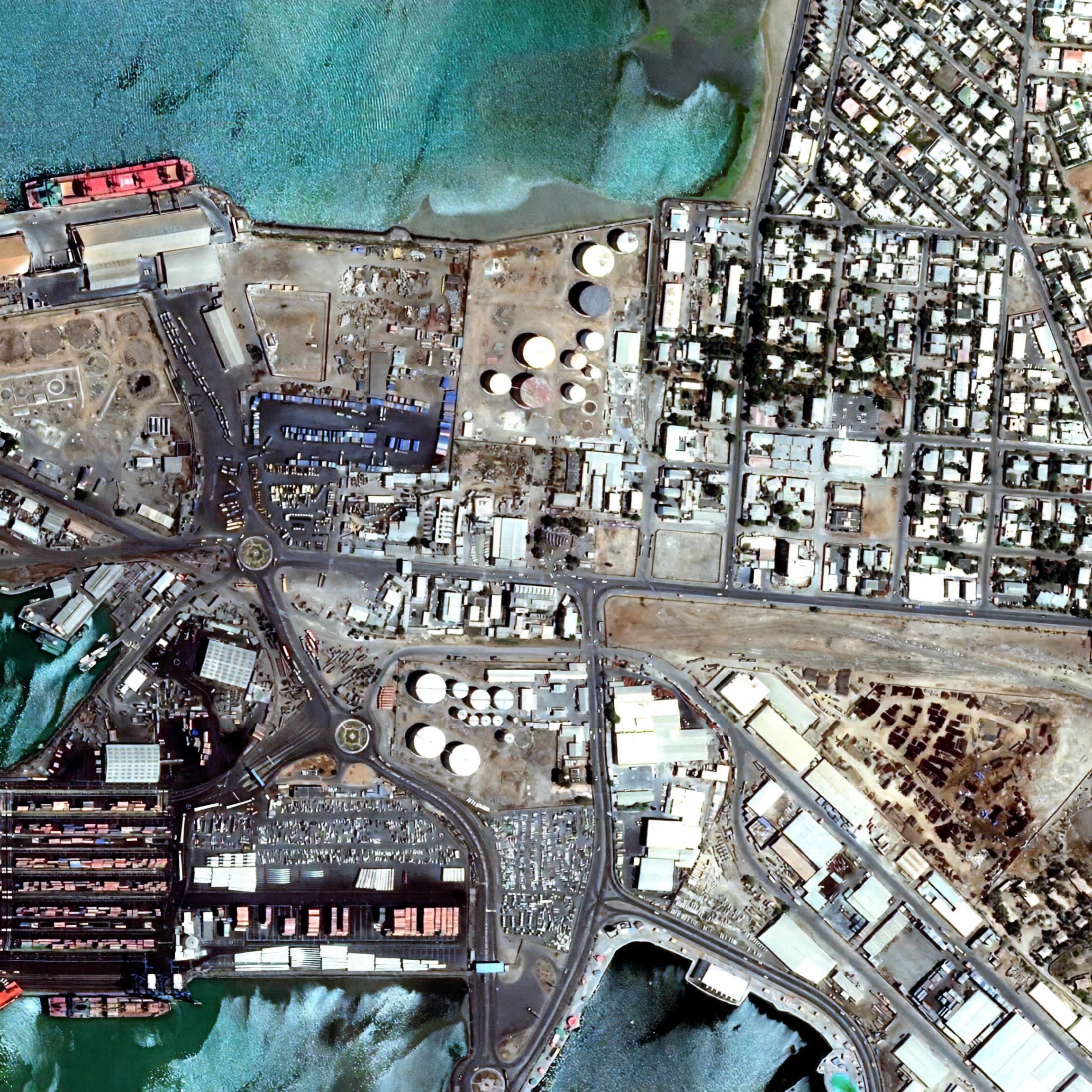}
    \caption{\textbf{Low bpp}: COLIC, PSNR=23.77, bpp=0.24.}
    \label{fig:colic}
\end{figure}

\begin{figure}[!ht]
    \centering
    \includegraphics[width=0.7\linewidth]{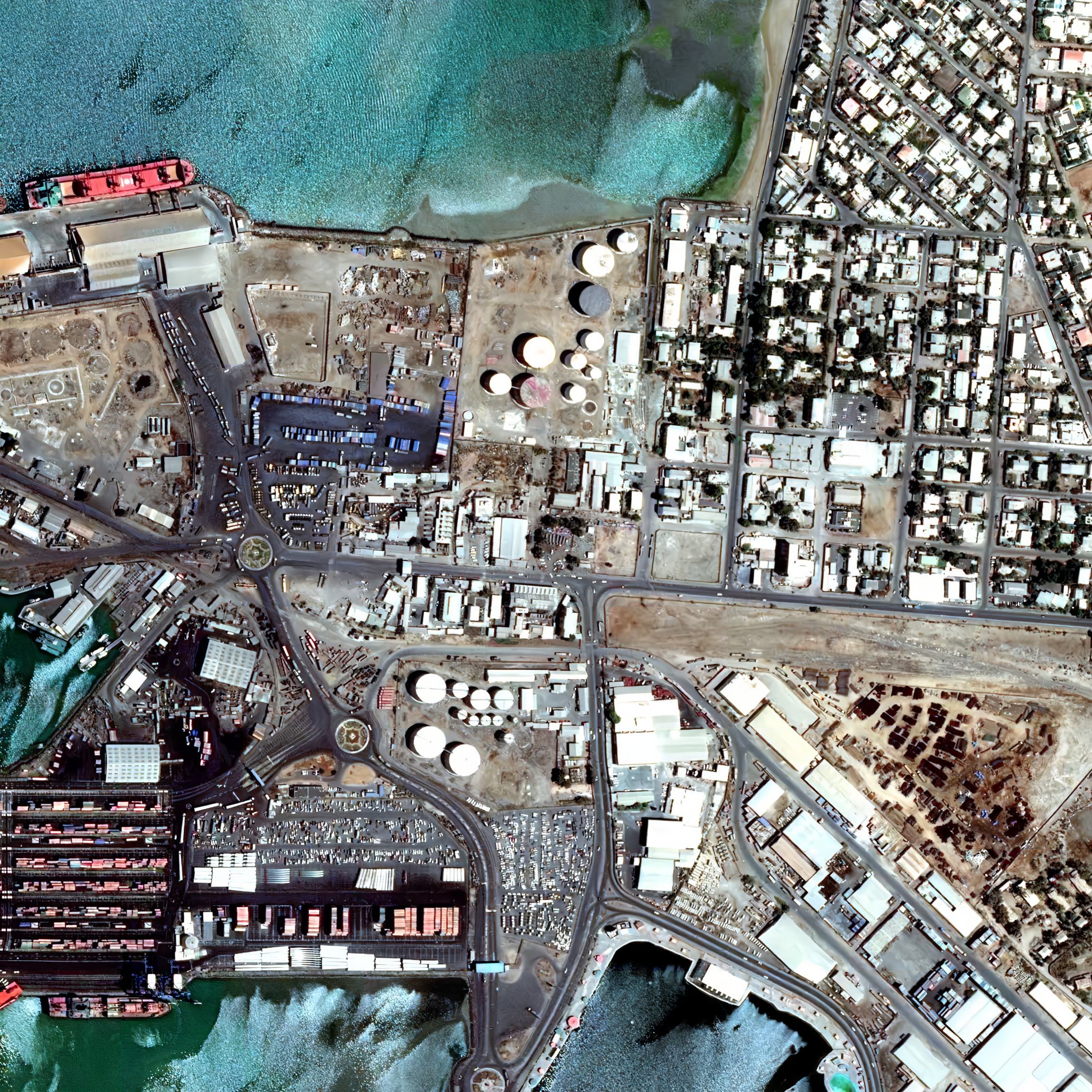}
    \caption{\textbf{Low bpp}: CDC, PSNR=24.82, bpp=0.41.}
    \label{fig:cdc}
\end{figure}

\begin{figure}[!ht]
    \centering
    \includegraphics[width=0.7\linewidth]{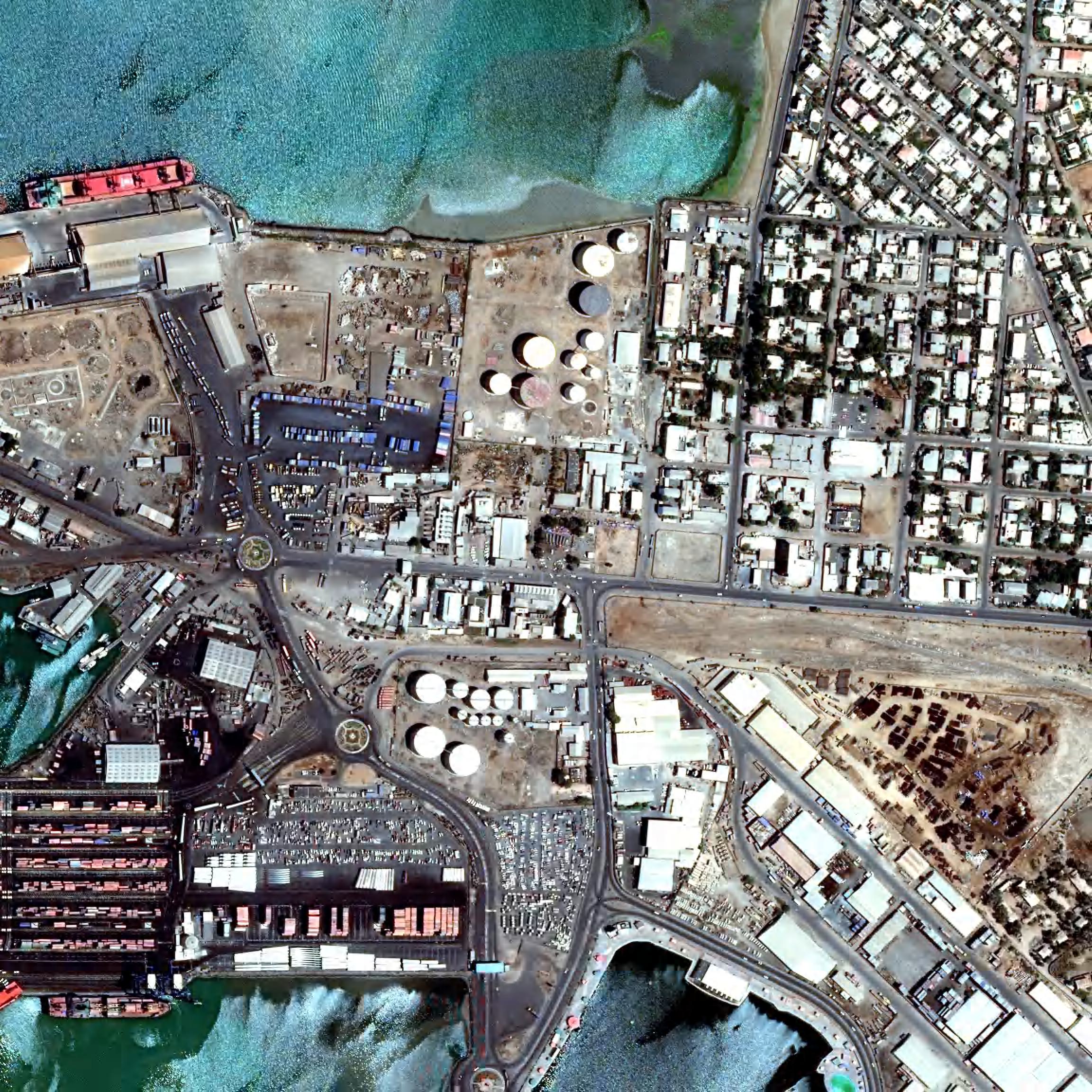}
    \caption{\textbf{Low bpp}: JPEG2000, PSNR=24.41, bpp=0.38.}
    \label{fig:jpeg2000}
\end{figure}

\begin{figure}[!ht]
    \centering
    \includegraphics[width=0.7\linewidth]{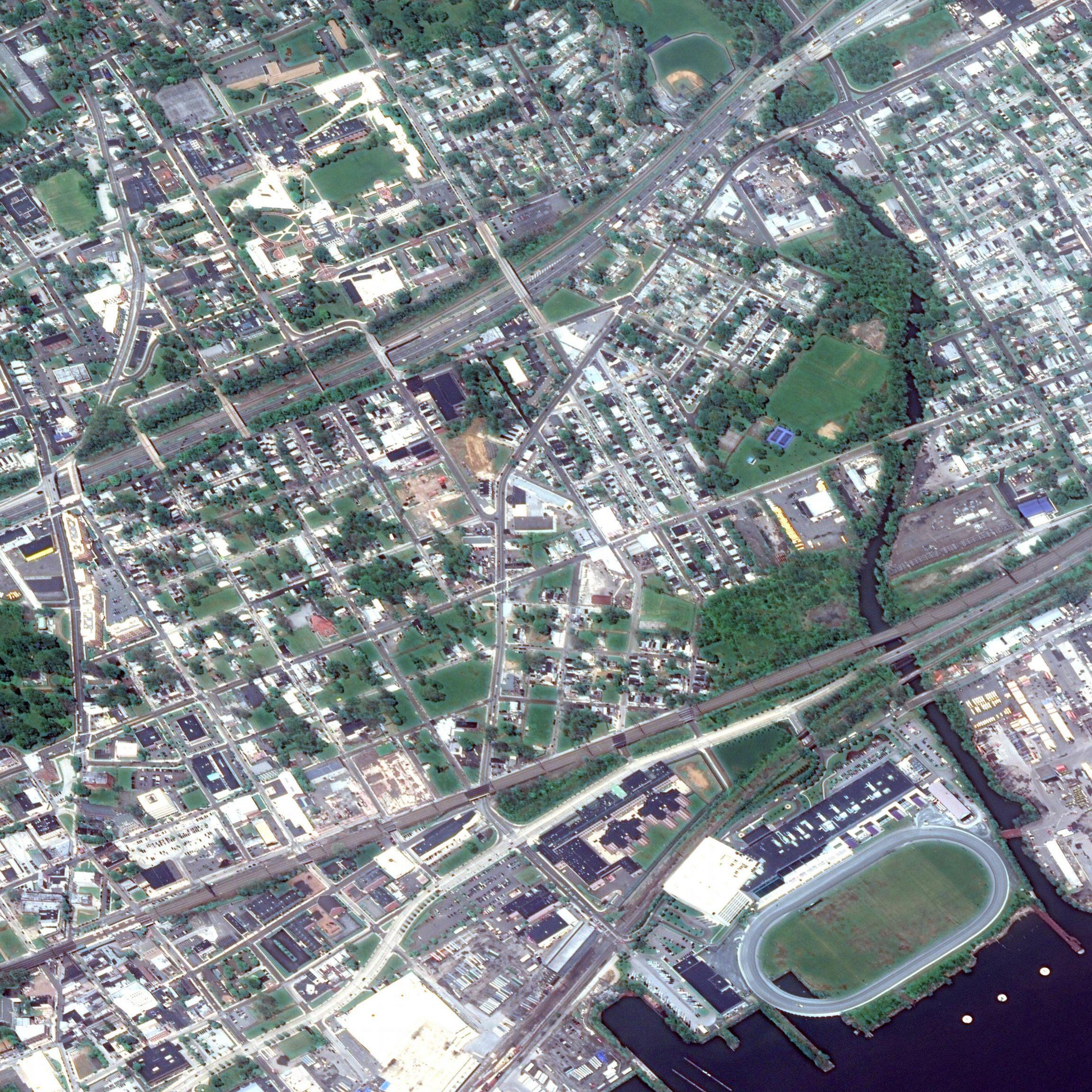}
    \caption{Ground truth.}
    \label{fig:gt2}
\end{figure}

\begin{figure}[!ht]
    \centering
    \includegraphics[width=0.7\linewidth]{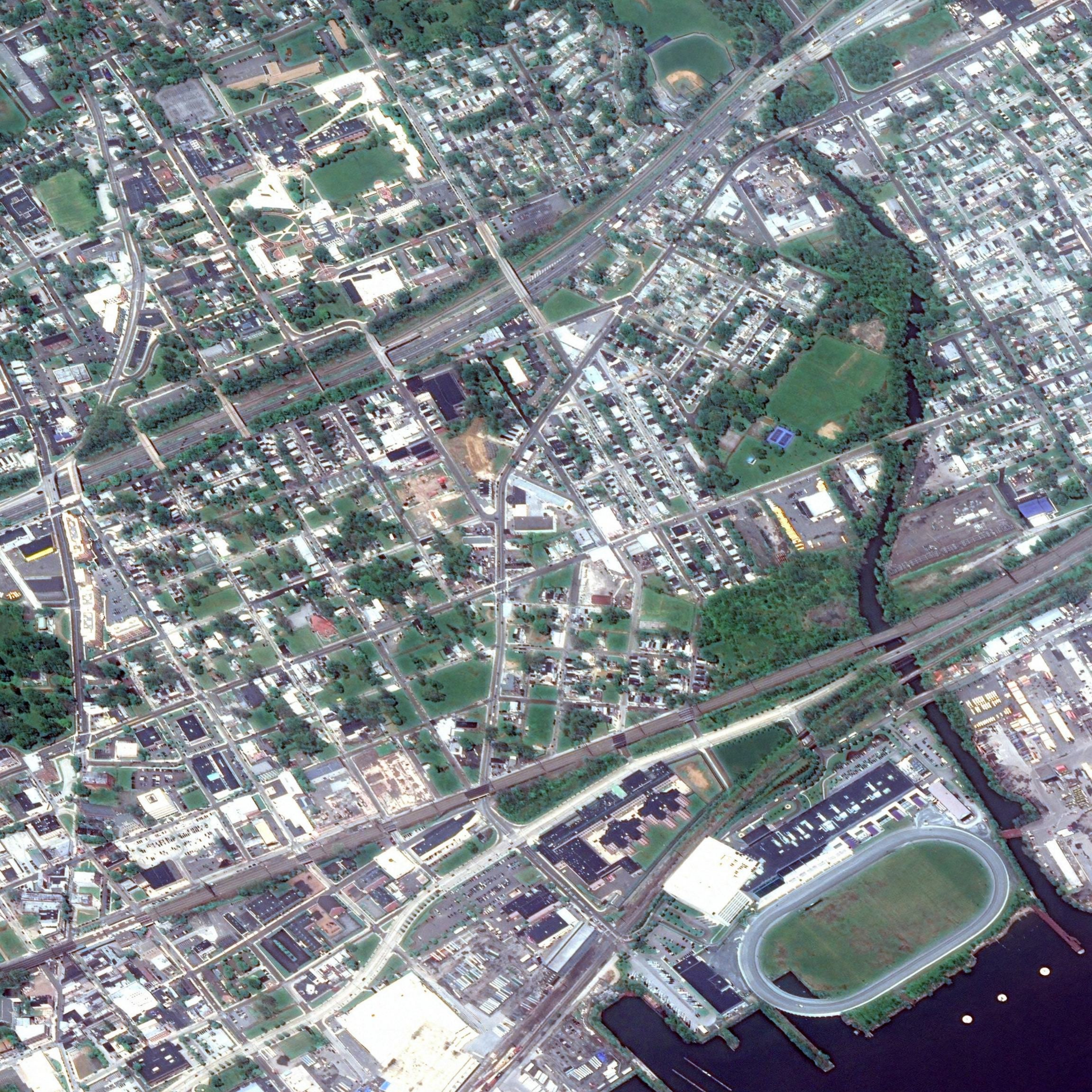}
    \caption{\textbf{High bpp}: \sysname, PSNR=30.73, bpp=0.67.}
    \label{fig:our2}
\end{figure}

\begin{figure}[!ht]
    \centering
    \includegraphics[width=0.7\linewidth]{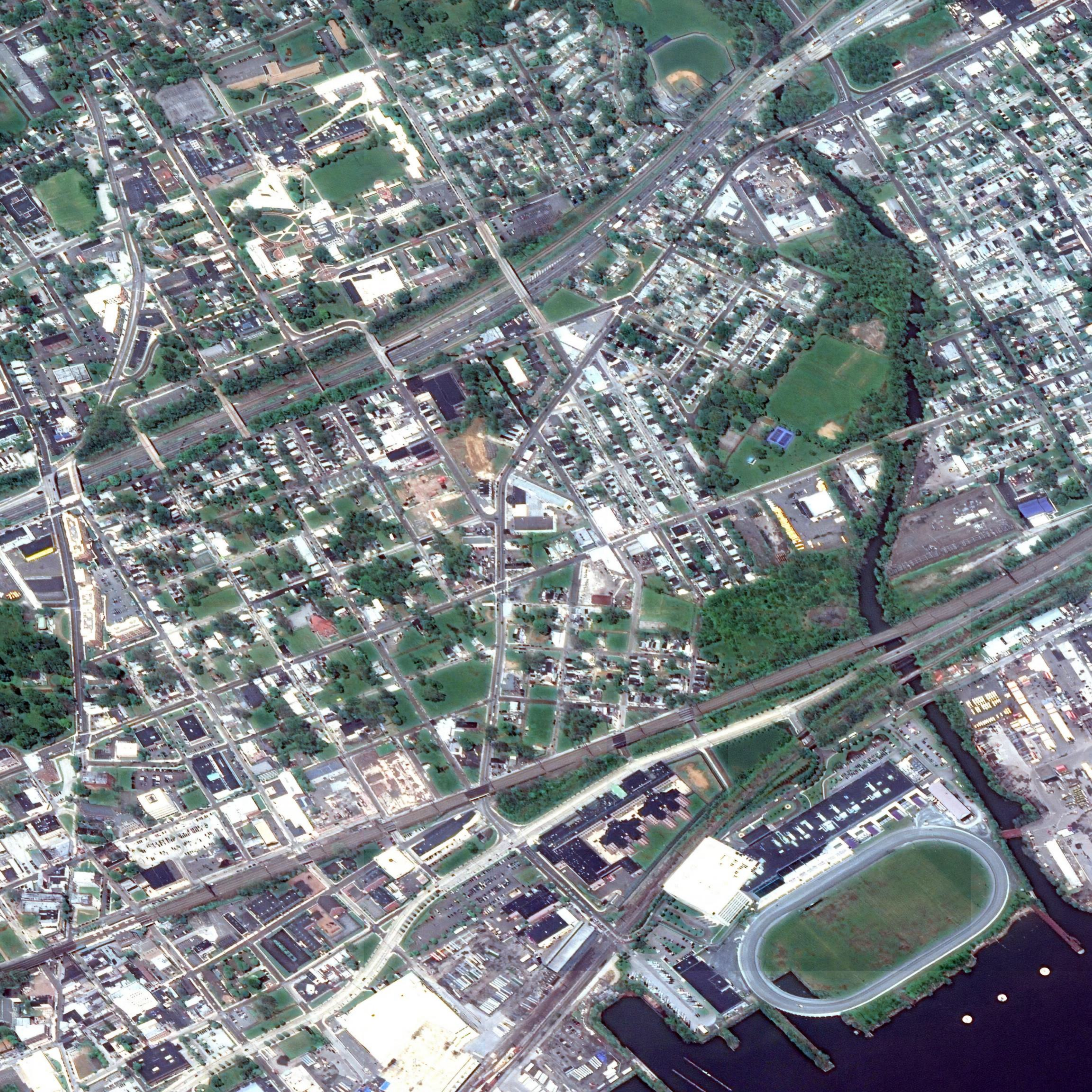}
    \caption{\textbf{High bpp}: HIFIC, PSNR=29.77, bpp=0.65.}
    \label{fig:hific2}
\end{figure}

\begin{figure}[!ht]
    \centering
    \includegraphics[width=0.7\linewidth]{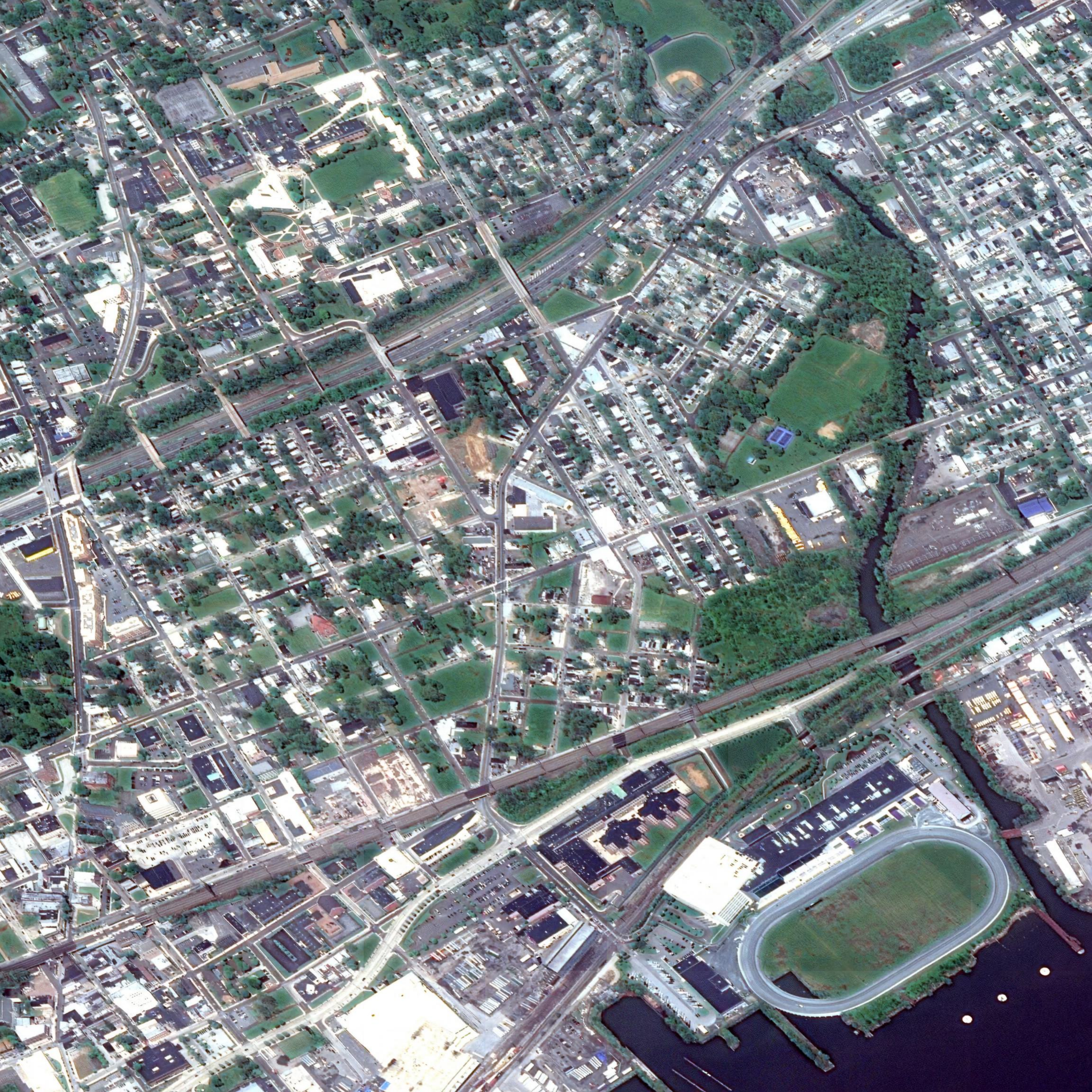}
    \caption{\textbf{High bpp}: COLIC, PSNR=29.82, bpp=0.70.}
    \label{fig:colic2}
\end{figure}

\begin{figure}[!ht]
    \centering
    \includegraphics[width=0.7\linewidth]{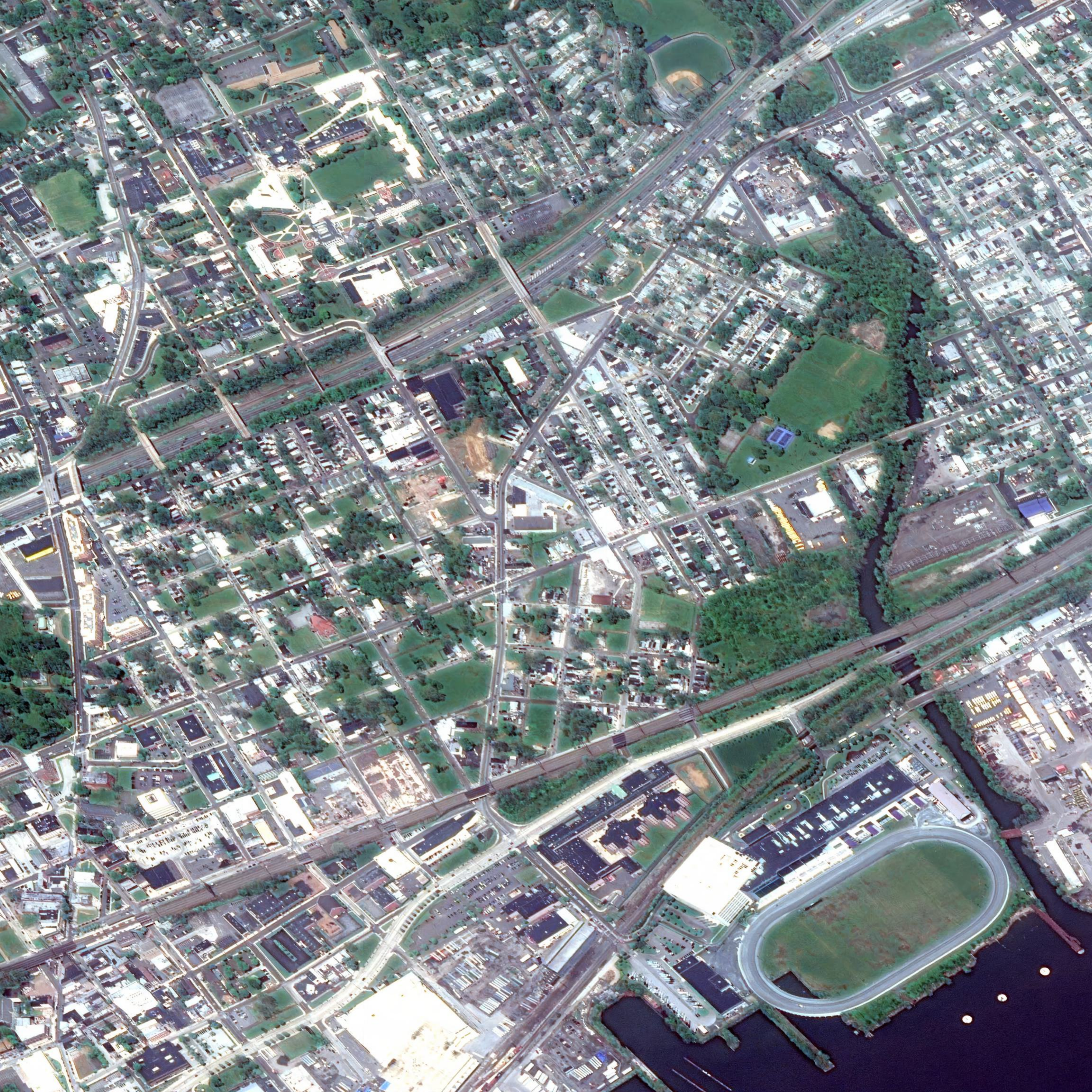}
    \caption{\textbf{High bpp}: CDC, PSNR=30.21, bpp=0.75.}
    \label{fig:cdc2}
\end{figure}

\end{document}